\documentstyle[12pt,epsfig]{article}
\begin{document}
\setlength{\baselineskip}{0.30in}
\newcommand{\beq}{\begin{equation}}
\newcommand{\eeq}{\end{equation}}
\newcommand{\be}{\begin{eqnarray}}
\newcommand{\ee}{\end{eqnarray}}
\newcommand{\nm}{\nu_\mu}
\newcommand{\nue}{\nu_e}
\newcommand{\nt}{\nu_\tau}
\newcommand{\nnt}{n_{\nu_\tau}}
\newcommand{\rnt}{\rho_{\nu_\tau}}
\newcommand{\mnt}{m_{\nu_\tau}}
\newcommand{\tnt}{\tau_{\nu_\tau}}
\newcommand{\bi}{\bibitem}
\newcommand{\ra}{\rightarrow}
\newcommand{\la}{\leftarrow}
\newcommand{\lra}{\leftrightarrow}

\begin{small}
{\hbox to\hsize{ \hfill TAC-1998-028}}
{\hbox to\hsize{October, 1998 \hfill  FTUV/98-68}}
{\hbox to\hsize{ \hfill  IFIC/98-69}}
\end{small}
\begin{center}
\vglue .06in
{\Large \bf {Unstable  massive tau-neutrinos and primordial nucleosynthesis.}}
\bigskip
\\{\bf A.D. Dolgov
\footnote{Also: ITEP, Bol. Cheremushkinskaya 25, Moscow 113259, Russia.}
\footnote{e-mail: {\tt dolgov@tac.dk}}, 
S.H. Hansen\footnote{e-mail: {\tt sthansen@tac.dk}}
 \\[.05in]
{\it{Teoretisk Astrofysik Center\\
 Juliane Maries Vej 30, DK-2100, Copenhagen, Denmark}
}}
\\{\bf S. Pastor \footnote{e-mail: {\tt Sergio.Pastor@uv.es}}} \\
{\it{Instituto de F\'{\i}sica Corpuscular - C.S.I.C.\\
Departament de F\'{\i}sica Te\`orica, Universitat de Val\`encia\\
46100 Burjassot, Val\`encia, Spain}}
\\{\bf D.V. Semikoz \footnote{e-mail: {\tt semikoz@ms2.inr.ac.ru}}} \\
{\it{Institute of Nuclear Research of the Russian Academy of Sciences\\
 60th October Anniversary Prospect 7a , Moscow 117312, Russia}}
\\[.40in]

\end{center}
\begin{abstract}
The impact of unstable Majorana $\nt$ on primordial nucleosynthesis is 
considered. The mass and lifetime of $\nt$ are taken in the
intervals 0.1-20 MeV and 0.001-400 sec respectively. 
The studied decay modes are
$\nt \rightarrow \nm +\phi$ and $\nt \rightarrow \nue +\phi$, where $\phi$
is a massless (or light) scalar. Integro-differential kinetic 
equations are solved numerically without any simplifying assumptions. Our
results deviate rather strongly from earlier calculations. Depending on 
mass, lifetime, and decay channels of the $\nt$, the number of effective
neutrino species (found from $^4$He), in addition to the 3 standard ones,
varies from -2 to +2.5. The abundances of $^2$H and $^7$Li are also
calculated.
\end{abstract}

PACS: 13.15.+g, 13.15.-r, 14.60.Pq, 98.80.Ft

Keywords: massive decaying neutrino, nucleosynthesis
\newpage

\section{Introduction}

Direct experimental data permits the tau-neutrino to be quite heavy,
$\mnt < 18 $ MeV~\cite{mnutau}, and if the mass indeed lays in the MeV
region, the $\nt$ would have a very strong impact on primordial
nucleosynthesis.  The recent data from Super Kamiokande~\cite{sk},
however, indicates possible oscillations between $\nm$ and either
$\nt$ or a new sterile neutrino, $\nu_s$, with a large mixing angle,
$\sin 2\theta \approx 1$ and a small mass difference $ \delta m^2 =
10^{-2} - 10^{-3}~\mbox{eV}^2$. If oscillations proceed into $\nt$
then both $\nm$ and $\nt$ are light and the impact of a $\nt$ mass on
nucleosynthesis is negligible. On the other hand it is not excluded
that the oscillations go into a sterile $\nu_s$ and, if this is the
case, $\nt$ may be heavy. Of course a heavy tau-neutrino should be
unstable because otherwise it violates Gerstein-Zeldovich
bound~\cite{gz}, and would overclose the universe.

If the life-time of $\nt$ is larger than a few hundred seconds, it can
be considered as effectively stable on the nucleosynthesis time scale.
The role of massive ``stable'' $\nt$ in the primordial nucleosynthesis
was considered in several papers with chronologically improving
accuracy. In pioneering work~\cite{ktcs} and a subsequent
paper~\cite{dr} the assumptions of kinetic equilibrium for all
participating particles and of validity of Maxwell-Boltzmann
statistics were made (see also the later paper~\cite{rrw}), hence the
problem was reduced to the solution of an ordinary differential
equation of Riccatti type. However, nonequilibrium corrections to the
spectra of both $\nt$ and massless neutrinos in the case of $\mnt$ in
the MeV range happened to be quite significant~\cite{dpv,ad} and a
more refined treatment of the problem had to be developed. In
ref.~\cite{fko} the simplifying assumption of Maxwell-Boltzmann
statistics was dropped in favor of the exact Fermi-Dirac one, but it
was assumed that kinetic equilibrium is maintained for all the
species.  Nonequilibrium corrections have been treated by one of the
authors~\cite{kk}, in the update to~\cite{fko}, who found that they do
not strongly change their original results. Exact numerical solutions
of the full system of kinetic equations for all neutrino species
without any simplifications have been done in refs.~\cite{hm,dhs}. In
the last work a somewhat better precision was achieved and in
particular a higher cut-off in particle momenta was used. Also the
expressions for some matrix elements of the weak interaction reactions
with Majorana particles were corrected. It has indeed been proven,
that nonequilibrium effects are quite significant, almost up to 50\%.

Consideration of the impact of an unstable $\nt$ on primordial
nucleosynthesis also has a long history. In the earlier papers [13-18]
%\cite{ks,rs,ts,dk,st,jm} 
some relevant effects were approximately estimated, but the accuracy
of the calculations was typically rather low. In the next generation
of papers [19-23]
%~\cite{kks,ksk,kkk,dgt,gt} 
the level of calculations was improved considerably but is still
subject to some criticism. General shortcomings of these calculations
are the assumption of kinetic equilibrium for some of the species or
the use of Maxwell-Boltzmann statistics. The recent
paper~\cite{kksato} makes use of similar simplifications. However, the
nonequilibrium corrections, as we have seen already for the stable
case and will see below for unstable $\nt$, are quite significant. The
corrections related to quantum statistics are typically at the level
of 10\%~\cite{dkai}.  The complete set of kinetic equations for the
case of the decay $\nt \rightarrow \nue + \phi$, where $\phi$ is a
massless scalar was numerically solved without any simplifying
assumptions in ref.~\cite{sh} for $\nt$ mass in the interval 0.1-1 MeV
and life-times larger than 0.1 sec.

In this paper we extend the analysis of ref.~\cite{sh} to a much
larger mass interval, $0.1 < \mnt/\mbox{MeV} < 20 $, and life-times between
0.001 and 400 sec both for the decays of Majorana type tau-neutrino
through the channels $\nt \rightarrow \nue + \phi$ and $\nt
\rightarrow \nm + \phi$.  Our numerical procedure is somewhat more
accurate than that of ref.~\cite{hm}, we have a larger cut-off in
dimensionless momentum $y=pa$, 20 instead of $ \sim 13$ ($a$ is the
expansion parameter of the universe), and we
correct some expressions for the matrix elements squared, as in our
previous paper related to the stable $\nt$. As in the preceding papers
on decaying $\nt$ we neglect all other interactions of the scalar
$\phi$, except for the decay and inverse decay. The exact treatment of
the scattering $\phi \nu \leftrightarrow \phi \nu$ and the
annihilation $ \bar\nu \nu \leftrightarrow 2\phi $ presents
considerable technical difficulties because of a larger number of
contributions into the collision integral and because of the
non-polynomial form of the matrix elements squared of these
processes. It will be considered in the subsequent
publication~\cite{dhps}.

The paper is organized as follows. After an overview in 
section~\ref{decays} we describe in detail our numerical solution
in section~\ref{numerical}. The results are presented in 
section~\ref{results} followed by a general discussion and 
conclusion in section~\ref{role}. Throughout the paper the 
natural system of units, $ \hbar = c = k = 1$, is used.

\section{Decays of tau neutrino}
\label{decays}

We assume that the $\nt$ is a Majorana type fermion which is coupled to 
a scalar boson $\phi$, possibly a Majoron~\cite{maj1,maj2}, 
which is light or even 
massless. The coupling of $\phi$ to neutrinos may have diagonal terms as e.g.
$g_1\bar \nt \nt \phi $ which are important for elastic scattering 
$\nt + \phi \leftrightarrow  \nt + \phi$ and annihilation
$\bar \nt + \nt \leftrightarrow  2\phi$. The non-diagonal coupling
$g_a\bar \nt \nu_a \phi $ is responsible for the decay of $\nt$ into lighter
neutrinos, $\nue$ or $\nm$ (correspondingly $a=e$ or $\mu$). 
We assume
that one of these two couplings dominates, i.e. that $\nt$ predominantly
decays either into $\nue \phi$ or $\nm \phi$ and consider these two 
possibilities separately. It is implicitly assumed that both $\nue$ and
$\nm$ are the usual active neutrinos. Since chirality is changed by the 
coupling to a scalar field, the corresponding light neutrinos should also 
be Majorana particles, otherwise new sterile states would be produced by the
decay.
We also assume, that $\phi$ is a weak singlet, because
the LEP measurements~\cite{z0} of the total decay 
width of $Z^0$ do not leave room for any other light weakly interacting 
particles except for the already known ones.

There are several possible ways of production of $\phi$ in the
primeval plasma.  The first and evident one is by the decay $\nt
\rightarrow \phi + \nu_a$.  Another
possibility is the annihilation $\nt+\nt \rightarrow \phi+\phi$ and
the third one is a possible non-thermal production in the course of a
phase transition similar to the production of axions at the QCD phase
transition. We neglect the last possibility, assuming that even if
(pseudo)goldstone bosons were created in the course of the phase
transition, the phase transition took place early enough so that the
created bosons were diluted by a subsequent entropy release in the
course of the universe cooling down. The rate of $\phi$-production in
$\nt$-annihilation can be estimated as:
\be
{\dot n_\phi \over n_{\nt} } = \sigma_{ann} v  n_{\nt} ,
\label{dotnphi}
\ee
where $v$ is the relative velocity and $ \sigma_{ann}$ is 
the annihilation cross-section. In the limit of large energies, 
$s=4E^2_{cm} \gg \mnt^2$ it is equal to:
$\sigma_{ann} \approx (g_1^4 /32\pi s)\ln (s/m^2_{\nt} )$
(see e.g.~\cite{dprv}). One can 
check that this rate is small in comparison with the universe expansion 
rate $H=\dot a /a$, if
$g_1 <10^{-5}$. In this case the production of Majorons through annihilation
can be neglected and they would dominantly be produced by the decay of $\nt$.
The opposite case of dominant production of $\phi$'s by $\nt$-annihilation 
and their influence on nucleosynthesis was 
approximately considered in ref.~\cite{dprv}.

The life-time of $\nt$ with respect to the decay into massless 
particles $\phi$ and $\nu_a$ is equal to:
\be
\tau_{\nt} = {8\pi \over g^2_a \mnt }.
\label{tau}
\ee
This is faster than the universe expansion rate if 
$T<4\cdot 10^{10} g_a\sqrt \mnt $, where the temperature $T$ and $\mnt$ are
expressed in MeV. Hence for $g_a > 10^{-10}$ the decay is essential,
i.e. $\tau_{\nu_\tau} < H^{-1}$, while
still $T>\mnt$. The interval of life-times of $\nt$, that we consider below,
$\tau_{\nt} = 0.001-400$ sec, corresponds to 
$g_a = \left( 4\cdot 10^{-9} - 6.5\cdot 10^{-12}\right)\sqrt {m/\mbox{MeV}}$.
Thus there is a large range of parameters (coupling constants and masses)
for which the decay is faster than the expansion while annihilation is
effectively frozen.

\section{Numerical solution}
\label{numerical}

\subsection{Kinetic equations}

The basic equations governing the evolution of the distribution
functions $f_a$ ($a=\nue$, $\nm$, $\nt$, and $\phi$) are discussed in
some detail in our previous works~\cite{dhs, dhs1}. The tables of
matrix elements squared of the relevant weak interaction processes, $\mid
A \mid ^2$, are presented there. We have some disagreement of our
expressions for $\mid A \mid ^2$ with those of refs.~\cite{hm,sh}. In
addition to the processes considered for the case of a stable $\nt$ we
should also take decay and inverse decay of $\nt$ into account and
include an extra distribution function, $f_\phi$. The distributions of
photons and electrons/positrons are assumed to have the equilibrium
form:
\be 
f_{e,\gamma} = \left[ 1 \pm \exp \left(E/T_\gamma \right) \right]^{-1},
\label{feq}
\ee
because the electromagnetic interactions are very strong and keep the
electromagnetic plasma in thermal
equilibrium. The behaviour of the unknown function of one variable $T(t)$ is
found from the law of covariant energy conservation:
\be
\dot \rho = -3H (\rho +P)~,
\label{dotrho}
\ee
where the Hubble parameter, $H=\dot a /a$, is expressed through the
total energy density, $\rho$, as $H^2 = (8\pi \rho / 3 m_{Pl}^2)$;
here $a(t)$ is the universe expansion factor (scale factor), P is the
total pressure and
$m_{Pl} = 1.22 \cdot 10^{19}$ GeV is the Planck mass.

The set of integro-differential kinetic equations has the form:
\beq{
 (\partial_t - Hp_j\partial_{p_j}) f_j (p_j,\,t) = I^{scat}_{j}+I^{decay}_{j},
\label{dtf1}
}\eeq
where the collision integral for two-body reactions
$1+2 \rightarrow 3+4$ is given by the expression:
\begin{eqnarray}
I^{scat}_{1} = {1\over 2E_1}\sum \int {d^3 p_2 \over 2E_2 (2\pi)^3}
{d^3 p_3 \over 2E_3 (2\pi)^3}{d^3 p_4 \over 2E_4 (2\pi)^3}\,
S\, |A|^2_{12\rightarrow 34}
\nonumber \\
(2\pi)^4\delta^{(4)} (p_1+p_2-p_3-p_4) F_{scat}(f_1,f_2,f_3,f_4)\, ,
\label{icoll}
\end{eqnarray}
$F_{scat} = f_3 f_4 (1-f_1)(1-f_2)-f_1 f_2 (1-f_3)(1-f_4)$, $|A|^2$ is
the weak interaction amplitude squared summed over spins of all
particles, and $S$ is the symmetrization factor which includes $1/2$
from the averaging over the spin of the first particle (if necessary), 
$1/2!$ for each pair of
identical particles in initial and final states and the factor 2 if
there are 2 identical particles in the initial state; the summation is
done over all possible sets of leptons 2, 3, and 4.

The "decay" part of the collision integrals enters the r.h.s. of the
equations for $f_{\nt}$, $f_{\nu_a}$ ($a=e$ or $\mu$), and $f_\phi$
and has respectively the form:
\be
I^{decay}_{\nt} = -{ m \over E_{\nt} p_{\nt} \tnt }
\int_{\left(E_{\nt}+p_{\nt}\right) /2}^{\left(E_{\nt}-p_{\nt}\right) /2} 
dE_\phi F_{dec} (E_{\nt},\, E_\phi,\, E_{\nt}-E_\phi),
\label{decnt}
\ee
\be
I^{decay}_{\nu_a} = { m \over E_{\nu_a} p_{\nu_a} \tnt}
\int^\infty_{\mid\left( m^2/4p_{\nu_a}\right) -p_{\nu_a} \mid} 
{dp_{\nt}p_{\nt} \over E_{\nt} } 
F_{dec} (E_{\nt},\,E_{\nt}-E_{\nu_a} ,\, E_{\nu_a} ),
\label{decnua}
\ee
\be
I^{decay}_{\phi} = {2  m \over E_{\phi} p_{\phi}\tnt }
\int^\infty_{\mid\left( m^2/4p_{\phi}\right) -p_{\phi} \mid}
{dp_{\nt}p_{\nt} \over E_{\nt} } 
F_{dec} (E_{\nt},\, E_\phi,\, E_{\nt}-E_{\phi} ),
\label{decphi}
\ee
where $m$ is the mass of $\nt$ (we omitted the index $\nt$) and:
\be
F_{dec} (E_{\nt},\, E_\phi,\, E_{\nu_a})=    
f_{\nt}(E_{\nt})\left[1+f_{\phi}(E_\phi)\right]
\left[1-f_{\nu_a}(E_{\nu_a})\right]  \nonumber \\
-f_{\phi}(E_\phi)f_{\nu_a}(E_{\nu_a}) \left[ 1- f_{\nt}(E_{\nt}) \right].
\label{fdec}
\ee
 
The contribution of the decay term, $I^{decay}$, into the collision integral 
of eq.~(\ref{dtf1}) is considerably simpler for numerical calculations than 
the contribution of scattering, $I^{scat}$,
because the former is only one-dimensional while the scattering terms 
only can be reduced to two-dimensional integrals.

Four integro-differential kinetic equations for the distribution
functions $f_j (p,t)$ $(j = \nt,\,\nm,\,\nue,\,\phi)$ and
eq.~(\ref{dotrho}) are solved numerically by the method described in
refs.~\cite{dhs,dhs1} (see also the earlier papers~\cite{semt}).
Technical details of the calculations are presented in the following
section.

It is convenient  instead of time and momenta to use
the following dimensionless
variables:
\beq{
 x= m a(t), \,\,  y_j= p_j a(t),
\label{xy}
}\eeq
where $m$ is an arbitrary parameter with dimension of mass,
which we took as $m=1$~MeV, and the
scale factor $a(t)$ is normalized so that $a(t) = 1/T_\nu = 1/T_\gamma$ at high
temperatures or at early times. In terms of these variables the kinetic
equations~(\ref{dtf1}) can be rewritten as:

\beq{
 Hx \partial_x f_j(x,y_j) = I^{scat}_j+ I^{dec}_j.
\label{hxdxf}
}\eeq

\subsection{Initial conditions}

For numerical integration of the kinetic equations we modified our
program for stable massive tau neutrinos (see the details of the
stable case calculation in our paper~\cite{dhs}) by including the
decay terms eq.~(\ref{decnt}-\ref{decphi}) in equations~(\ref{hxdxf}).

All the calculations of the two-body-reaction part of the collision
integral remain the same as in our previous work. So we discuss below
only the changes to the program related to the decay terms. We solve
the system of kinetic equations~(\ref{hxdxf}) in the 'time' interval
$x_{in} \le x \le x_{fin} $ under several assumptions concerning
decay.  First, we assume that the decay was negligible before the
initial time $x_{in}$ and take the initial majoron distribution
function equal to zero, $f_\phi(x_{in},y)=0$. We can always do this
for sufficiently small $x$, because the collision rates in
eq.~(\ref{hxdxf}) are very high due to the factor $1/x^2$ while the
decay terms are proportional to $x^2/\tau$ and even for very small
life-times, $\tau \ll 1$ sec they are suppressed at small $x$. To
determine the proper initial moment $x_{in}$ we approximately solved
the kinetic equation for $f_\phi (x,y)$ neglecting the contribution from
$\phi$ in the collision integral and assuming equilibrium
distributions for all other participating particles.  Then we can
integrate this equation over $\phi$ momentum and obtain the following
equation for the energy density of majorons at early times:
\beq{
 \partial_x (x^4\rho_\phi) = 2 C_\phi x^2
 \frac{\mnt}{\sqrt{x^4\rho_{tot}}\tau}
\frac{1}{2\pi^2}\int_0^\infty \frac{ydy}{e^y -1} 
\log\frac{1+\exp(\frac{\mnt^2 x^2}{4y})}
 {\exp(-y)+\exp(\frac{\mnt^2 x^2}{4y})}~,
\label{dener_phi}
}\eeq
where $C_\phi = 1.221 \cdot 6.582  \sqrt{3/(8\pi)}= 2.777$ is a constant 
and $x^4 \rho_{tot}=10.75 \pi^2/30$ is the total energy density of photons,
$e^{\pm}$, and the three neutrino species in comoving frame,
and the lifetime, $\tau$, is given in seconds.

For a small value of the product $\kappa=(\mnt^2 x^2/4) \le 1 $ 
we can solve this 
equation and obtain:
\begin{eqnarray}
\rho_\phi &=& 2 C_\phi \frac{x^3}{3x^4} \frac{\mnt}{\sqrt{\rho_{tot}}\tau} 
(\frac{\zeta(3)}{2\pi^2}-\frac{\log(2)}{24} )(1-0.08 \kappa
 + 0.006 \kappa^2 + O(\kappa^3)) 
\nonumber \\
&\approx& 0.31\frac{\mnt}{x\tau}  (1-0.08 \kappa
 + 0.006 \kappa^2 + O(\kappa^3))~.
\label{ener_phi}
\end{eqnarray}

From equation~(\ref{ener_phi}) we can find the energy density of
scalars for a given initial time $x_{in}$ and compare it with the
energy density of equilibrium massless neutrinos
$\rho_{eq}=0.57573/x^4$.  We will assume that scalars are absent if
$\rho_\phi/\rho_{eq}<0.01$.  For  $\nu_\tau$ lifetime in the range
$1<\tau/\mbox{sec}<\infty$ we take the initial time $x_{in}=0.1$ for
all values of the mass. For $0.1<\tau/\mbox{sec}<1$ we take
$x_{in}=0.1$ for $\nt$ mass in the range $0.1 < \mnt/\mbox{MeV} <2$
and $x_{in}=0.05$ for $2 < \mnt/\mbox{MeV} <20$. For $\tau/\mbox{sec}<0.1$
we always take  $x_{in}=0.05$.

\subsection{Choice of the final time}
 
The contributions to the collision integral in the r.h.s. of
eq.~(\ref{hxdxf}) from two-body reactions are suppressed at large $x$
by at least the factor $1/x^2$.  We find that at $x \approx 50$ all
the quantities, which do not feel the decay, reach their asymptotic
values so, to be on the safe side, we choose the final time for the
reactions $x_{scat} = 100$.  Beyond this time all collision terms are
taken to be zero.

The decay continues until all the  tau-neutrinos have  disappeared.
In our runs we assume that the decay terminates at $x_{decay}$ at which 
$\rho_{\nt}/\rho_{eq}<10^{-4}$,
where $\rho_{eq}=0.57573/x^4$ is the energy density of one massless 
equilibrium neutrino.
As the final time $x_{fin}$ we take the maximum of the two, 
$x_{decay}$ and $x_{scat}$. 
  
The nucleosynthesis code of ref.~\cite{kaw} requires 
the final time $x \approx 2000$ for the total
energy density, photon temperature  and $n \leftrightarrow p$ rates. We applied 
a separate program,
which calculates these quantities, using the energy conservation law and
 our final values for the distribution 
functions $f_{\nu_i}$ and the temperature $T_\gamma$ at time $x_{fin}$. 

\subsection{Momentum grid}

The particles produced by $\nt$ decays have momenta around $y= x \mnt/2 $.
The decay becomes operative and starts to diminish the $\nt$ number 
density at the time  $t \approx \tnt$ or at $x_D \sim \sqrt{\tnt}$,
if the inverse decay is not 
important, i.e. if the mass and life-time are sufficiently large. 
Thus for the decay products we should take into account 
momenta up to $y_{D} = C \sqrt{\tnt} \mnt$, where the coefficient $C$  
depends on the time when the decay is completed.

On the other hand the distribution functions of the tau neutrino  
and of the massless neutrino, which does not participate in the decay, 
are suppressed at high momenta. Already at $y=20$ we have
$\rho(y>20)/\rho \sim 10^{-5}$ for these particles,
and we therefore choose the momentum cut-off equal 
to $y_{L}=20$. The cases $y_{D} > y_{L}$ and $y_{D} < y_{L}$ are 
considered separately.

When $y_{D} < y_{L}$ (it corresponds to $\tnt<0.1$ sec 
for $\mnt >1$ MeV and $\tnt<1$ sec for $\mnt<1$ MeV)
we take a logarithmic grid for  the Majorons in the 
region $0.01<y<20$ and a linear grid for the neutrinos in the region $0<y<20$.

Since the collision integrals for two-body reactions are
two-dimensional, their calculation takes most of the CPU time, and
correspondingly we confined ourselves to a 100 point grid for the
reaction terms.  On the other hand, the decay terms are
one-dimensional and we took a 1600 point grid in their calculation.

When $y_{D} > y_{L}$ (it corresponds to  $\tnt>0.1$ sec 
for $\mnt >1$ MeV and $\tnt>1$ sec for $\mnt<1$ MeV)
we took a linear grid for the decay products in the interval $0<y<y_D$
while for the neutrino, which does not participate in the decay,
the linear grid was taken in the range  $0<y<20$.
In the interval  $0<y<20$ we took 100 points in the grid, and expanded with an
equally spaced grid with the same $dy$ for the decay products up to  
$y=y_D$. We  calculated the reaction terms in the collision integrals only 
in the interval $0<y<20$, because the decay into modes with $y>20$ becomes 
important when the contribution of reactions into the collision integral is 
already suppressed by the factor $1/x^4$ with $x\gg 1$ for any mass.

At $y=0$ the analytical expressions for the 
collision integral for $\nt$ differ from those at $y\neq 0$ due to 
$0/0$-uncertainty arising from the factor $1/E_1 p_1$ in front of the integrals
and similar vanishing factors in the D-functions (see their definition in our 
previous papers~\cite{dhs,dhs1}). Because of that, we consider the
point $y=0$ separately. This permits to compare the collision integrals 
at $y=0$ with those in nearby points, $y \ll 1$, in order to
check that numerical errors  are small in the region of small momentum,
$y<1$. This is important for a massive tau-neutrino, because its distribution 
function rapidly changes in the course of evolution.  

%\subsection{Time evolution}
%
%Evolution in time is calculated by the Euler method.
%We checked that numerical errors in the distribution functions calculated by
%the Euler method are smaller than $0.1\%$. The Euler method
%allows us to save the factor of 2 in CPU time compared 
%to the second order Runge-Kutta method and more than the factor of 10  
%compared to the Bulirsch-Stoer method (see e.g.~\cite{numrec}).

An important check of our program is the comparison of the run with the 
large life-time, $\tnt=400$ sec, with that with an infinite life-time, 
$\tnt=\infty$, (when the decay terms in the collision integral are switched 
off) for masses in the interval $0.1 < \mnt/\mbox{MeV} <20$. 
The results of both these runs are   very close  for all masses.

\subsection{Nucleosynthesis}
\label{nucl}
In order to extract the implications of a massive decaying $\nt$ on
light element abundances, we have modified the standard
nucleosynthesis code (ref.~\cite{kaw}) in the following way.  First we
import the final photon temperature, which is different from
$1.401/x$, in order to specify the final value of the baryon to
photon ratio, $\eta_{10} = 10^{10} n_B/n_\gamma$.  We
calculate various quantities at the correct photon temperature and
import the values to the code.  These imported quantities are the
total energy density $\rho_{tot}$, the 6 weak interaction rates for
$(n \leftrightarrow p)$-reactions, and finally $d (\ln \, a^3) /d
T_\gamma$ (with the account of neutrinos) which governs the evolution
of the photon temperature. The excessive number of equivalent
neutrinos, $\Delta N$, which may be both positive and negative, 
is interpolated from a look-up table as a
function of $^4$He and $\eta_{10}$.

\section{Results}
\label{results}

The impact on nucleosynthesis of a heavy decaying tau neutrino strongly 
depends upon the decay channel. In the case of the decay into $\nm +\phi$
the most important effect is the overall change in the total energy density
and the corresponding change of the universe cooling rate. Nonequilibrium
corrections to the spectra of $\nue$ are relatively weak for a small 
life-time,
such that practically all $\nt$ have already
decayed at the moment of neutron-proton 
freezing, $T\approx 0.5 $ MeV. For a larger life-time some nonequilibrium
$\nue$ would come from annihilation $\nt +\nt \rightarrow \bar \nue + \nue$ 
and, as is well known, would directly change the frozen $n/p$-ratio. The
distortion of the $\nue$ spectrum is much stronger in the case of the decay
$\nt \rightarrow \nue +\phi$. Correspondingly the results for the light
element abundances are considerably different for these two different decay
channels.

In figs.~1 the total energy density of all three neutrino species and
of the scalar $\phi$ is presented as a function of the tau neutrino
mass and life-time for asymptotically large values of "time" $x$. It
is given in terms of the effective number of massless equilibrium
neutrinos,
$N_{eff}=(\rho_{\nt}+\rho_{\nt}+\rho_{\nt}+
\rho_{\phi})/\rho_{eq}$, where $\rho_{eq}= (7\pi^2/120 ) m_0^4/x^4$ is
the energy density of one massless neutrino. 
To avoid confusion we should stress, that this effective
number of neutrinos, $N_{eff}$, has nothing to do with the excessive
number of equivalent neutrinos, $\Delta N$, which is introduced to
describe the change in the primordial $^4$He abundance (see
below). The behavior of the "iso-neutrino" curves is quite
transparent: for a large life-time the relative contribution from
nonrelativistic $\nt$ into $\rho$ increases as the first power of the
scale factor $a\sim x$. For larger masses the $\nt$ becomes
nonrelativistic earlier, and its contribution into the energy density
is bigger.  On the other hand, very heavy tau-neutrinos annihilate more
efficiently and thus their frozen number should be smaller than the
number density of less massive ones.  This explains the extrema in
fig.~1a.  
For small masses and lifetimes a plateau is reached, where a fraction
of percent uncertainty makes the iso-curves wiggle. This uncertainty 
arises from merging of different programs for different mass-lifetime
regions.
Somewhat less trivial is the fact, that for larger values of
mass and relatively small life-times the energy density can be smaller
than in the standard model. This part of the graph is enlarged and
presented in fig.~1b. A decline in the energy density in this range is
related to the early decay of $\nt$, when its number density is not
yet frozen with respect to annihilation and is Boltzmann suppressed.

In figs.~2 and 3 the time evolution of the energy (a) and number (b)
densities of different species are presented. The behavior of $\nm$
and $\phi$ reflect the behavior of $\nt$. For a small mass and
life-time (fig.~2) $\nt$ remains relativistic and its number and
energy densities follow the equilibrium massless ones. For $x> 1$,
when the decay starts to operate, some decrease in $n_{\nt}$ is
observed due to filling of the previously empty scalar states. However,
the energy density, $\rho_{\nt}$, does not decrease starting from
$x\approx 5$ due to the above mentioned  effect of domination of
nonrelativistic particles.  Later, at $x>20$, the Boltzmann suppression
becomes essential and both $n_{\nt}$ and $\rho_{\nt}$ quickly drop
down. For large mass and life-time (fig.~3) the features in the picture are
more pronounced. The decrease in $n_{\nt}$ and $\rho_{\nt}$ for
$0.1<x<0.5$ is related to the Boltzmann suppression of almost equilibrium
$\nt$. After that, the  $\nt$-annihilation becomes frozen and the energy
density goes up. Above $x=3$ the decay comes into play and $\nt$
disappear from the primeval plasma. To test the consistency of our
calculations we checked the conservation of particle number densities
and found it valid with an accuracy better than 1\%.

Our main results for the decay channel $\nt \rightarrow \nm + \phi$
are presented in figs.~4. We calculated the production of $^4$He for
different masses and life-times, using the nucleosynthesis code
modified as described in section~\ref{nucl}, and expressed the
variation of the mass fraction of $^4$He in terms of the extra number of
neutrino species which gives rise to the same amount of $^4$He. The
results are presented in two graphical forms: as a three dimensional
graph $\Delta N = F (\mnt,\,\tau_{\nt})$ (fig.~4a) and as a series of
curves $\Delta N = F (\mnt)$ for different values of
$\tau_{\nt}$ (fig.~4b). For large masses and low life-times $\Delta N $ is
negative. This is related to the decrease of the energy density
discussed above. Thus if $\mnt = 10$ MeV and $\tau_{\nt} = 0.1$ sec,
the effective number of neutrino species at nucleosynthesis is only
2.5.

Comparison with the results of other groups shows a rather strong deviation.
We ascribe this to the simplifying approximations made in the earlier papers, 
which might give rise to a significant difference with the exact 
calculations,
and to a better accuracy of our numerical calculations, which is typically 
at the fraction of per cent level. For example in the case of 
$\mnt = 14 $ MeV and $\tnt = 0.1 $ sec we obtain for the energy density
$\rho/ \rho^{eq}_{\nu_0} = 2.9$, while the group~\cite{kksato} obtained 2.5.
In the limit of small life-times and masses our results are close to 3.57 
(this is the energy density of three light neutrinos and one scalar), while
the results of~\cite{kksato} are close to 3.9. The effective number of
neutrino species found from $^4$He in our case is 
$3+\Delta N=2.9$ for $\mnt = 10$ MeV and
$\tnt = 1 $ sec, while that found in ref.~\cite{kkk} is 3.1. The difference is
also large for $\mnt = 10$ MeV and $\tnt = 0.01 $ sec: we find 
$3+\Delta N=2.66$
and the authors of~\cite{ksk} obtained 2.86.

Consideration of the decay $\nt \rightarrow \nue + \phi$ proceeds
essentially along the same lines. The energy densities of different
particle species in this case are almost the same as in the case of
the decay $\nt \rightarrow \nm + \phi$, so we do not present
them. However, the influence on light element abundances is now
very different from the previous case, because the nonequilibrium
electronic neutrinos produced by the decay strongly shift the $n/p$ ratio. In
particular an excess (with respect to the equilibrium distribution) of
low energy electronic neutrinos would create a smaller $n/p$-ratio and
thus a correspondingly smaller mass fraction of $^4$He.

The light element abundances for different values of $\mnt$, $\tnt$
and the baryon-to-photon ratio $\eta_{10} = 10^{10} n_B/n_\gamma$ are
summarized in figs.~5-11. Massive unstable $\nt$ may act both ways,
diminishing and increasing the mass fraction of $^4$He. In terms of
the effective number of neutrino species it may correspond to a large and
positive $\Delta N$, up to $\Delta N = 2.5$, or even to the negative
$\Delta N = -2$ (see fig.~7).
As seen on the $\Delta N$ iso-curves (figs.~8), there is a minimum
in $\Delta N$ for lifetime $\sim 0.05$ sec and mass $\sim 4$ MeV.

Also for this channel we disagree with previous papers. E.g. for
$m=0.6$ MeV and $\tau = 100$ sec we find $Y({^4 \mbox{He}}) \approx 0.244$,
whereas ref.~\cite{sh} obtains  $Y({^4 \mbox{He}}) \approx 0.20$.

It was shown in ref.~\cite{rs} that late decaying $\nt$ with $\tnt =
10^3-10^4$ sec and $\mnt > 3.6 $ MeV would distort the deuterium
abundance strongly if the decay proceed into electronic neutrinos.  These
$\nue$ would create excessive neutrons through the reaction $\nue + p
\rightarrow n + e^+$, which would form extra $^2$H. 
This is seen explicitly on fig.~9b, where $^2$H clearly increases
as a function of lifetime. The deuterium production goes quadratically
with the baryon density, and we see, that this effect is much less
pronounced for low $\eta_{10}$.

In figs.~9-11a the mass fraction of $^4$He is presented as a function of
$\mnt$ for several values of $\tnt$ for different baryon-to-photon ratio,
$\eta_{10} = 10^{10} n_B/n_\gamma = 3,5,7$. In figs.~9-11b,c the similar 
graphs for the primordial $^2$H and $^7$Li are presented.

More graphs showing various elements ($^2$H, $^4$He and $^7$Li) and as 
functions of mass and lifetime, for both channels 
$\nt \rightarrow \nu_\mu + \phi$ and
$\nt \rightarrow \nu_e + \phi$ and for $\eta_{10} = 1,3,5,7,9$ 
can be found on the web-page: 
{\tt http://tac.dk/\~{}sthansen/decay/}
together with plots of the n-p reaction rates.

Finally we changed the initial condition, allowing the majorons to 
be in equilibrium at $x_{in}$, $ f_\phi (x_{in}) = 1/(\exp(y)-1)$.
This situation corresponds to the majorons 
being  produced by some other mechanism prior to the $\nt$ decay
as discussed in section~2.
In this case the inverse decay is more efficient than for 
$f_\phi (x_{in}) = 0$ and $\rho_{\nt}$ 
decreases slower as is seen from fig.~12. 
The change in $\Delta N$, as compared to the case when 
$f_\phi (x_{in}) = 0$, varies between 0.4 and 1.0, 
$\Delta N _{f_\phi = eq} = \Delta N _{f_\phi = 0} + (0.4 - 1.0)$,
depending on mass
and lifetime. In particular for long lifetime this difference goes to
0.57 as expected for all masses.

\section{The role of possible neutrino oscillations}
\label{role}

The considerations of this paper neglect a possible influence of
neutrino oscillations on primordial nucleosynthesis. If the Super
Kamiokande result~\cite{sk} is true, then $\nm$ is strongly mixed with
another neutrino, either $\nt$ or a sterile one. If $\nm$ is mixed
with $\nt$, then the mass of the latter is small, $\mnt < 160 $ keV and
our results obtained for larger masses would be irrelevant. In the
other possible case, when oscillations are between $\nm$ and a sterile
one, $\nu_s$, the $\nu_\tau$ mass is not restricted other than by the
direct laboratory limit~\cite{mnutau} and our calculations make sense
and permit to improve the latter significantly. Still the oscillations
into a sterile neutrino state would have an important impact on 
nucleosynthesis and would change our results obtained without
oscillations. We postpone a detailed numerical calculations of the
influence of neutrino oscillations on primordial nucleosynthesis for
the future study and give only some simple estimates here (a brief 
discussion of the impact of the Super Kamiokande results on nucleosynthesis 
is also given in  ref.~\cite{kkss}).

In accordance with the measurement~\cite{sk} the mixing angle is
surprisingly large, $\sin^2 2\theta = 0.8 -1 $ and vacuum oscillations
are not suppressed at all. Interactions with the thermal bath generally
result in a suppression of oscillations, except for a possible
MSW-resonance. This has been studied in a large number of works and a
list of references can be found in the recent papers~\cite{bfv,chki,bvw}. It
is sufficient for our purposes to use the simple results of
the early papers~\cite{bd,ekt}, 
that the oscillations $\nm \leftrightarrow \nu_s$ would be
fully developed if $\sin^4 2\theta |\delta m^2 | \geq 3\cdot 10^{-3}
\mbox{eV}^2$~\cite{bd} and 
 $\sin^2 2\theta \cdot \delta m^2  >  3\cdot 10^{-6}
\mbox{eV}^2$~\cite{ekt}.
Thus for $|\delta m^2 | \leq 10 ^{-3}$ eV$^2$ the oscillations
$\nm\lra \nu_s$, in the case that $\nm$ were produced by the usual
weak interaction processes in the primeval plasma, would not
change our results considerably. The oscillations of muonic neutrinos
created in the $\nt$ decays would not influence nucleosynthesis if the
$\nt$ life-time is large enough. In this limit the oscillations do not
change the total energy density of  the plasma in comparison with the
no-oscillation case.  The decays with a short life-time and the
subsequent oscillations would somewhat amplify the production of extra
neutrino species in the primeval plasma and might be noticeable in
nucleosynthesis.

The oscillations might be very strongly suppressed if the cosmic lepton 
asymmetry (in the case considered it is muon asymmetry) is large in
comparison with the usually accepted value $10^{-9} - 10^{-10} $. The 
calculations made in this work should be correspondingly modified to take
into account the change in the distribution function, $f_{\nu_\mu}$, due to 
the non-vanishing muonic chemical potential. However, even an increase of muon
asymmetry by 6-7 orders of magnitude would have a negligible effect on 
the kinetics  of nucleosynthesis. This is not so for the oscillations of
$\nue \lra \nu_s$, when a large electric asymmetry might be generated by
the oscillations and could shift the equilibrium value of $n/p$-ratio
(for a recent work and the list of references see~\cite{bfv,chki,bvw}).

If one sterile neutrino species exists, then it is rather natural to expect
three sterile ones. This possibility could be realized for example if 
the mass matrix of neutrinos contains both Majorana and Dirac mass 
terms~\cite{adnu,pl}. In this case, if $\nt$ indeed is heavy and unstable,
the anomalies in neutrino physics should/could be explained by the 
oscillations in the sector $\nue - \nm - \nu_{s1} -\nu_{s2}$. Due to these
oscillations there might be effectively up to two extra neutrino species
at nucleosynthesis, but a massive and unstable $\nt$ which could effectively
create a negative $\Delta N$ would permit to avoid a contradiction with the
data.

For the case of neutrino interactions with majorons the contribution
into the neutrino refraction index from majorons in the medium would
strongly suppress oscillations~\cite{br}, if the $\nt \nm
\phi$-coupling constant is larger than $10^{-6}-10^{-7}$. It
corresponds to the very short life-time $\tau_{\nt} \approx 10^{-6}$
sec $(10^{-7} /g)^2\, (\mbox{MeV} / m_{\nt})$.

Thus to summarize, our calculations should be considerably modified if 
neutrino oscillations, as indicated by Super Kamiokande data, indeed exist.
However, first, it seems premature to make a strong conclusion 
based on these data and, second,
there exist regions in the parameter space for which our results with the 
neglect of oscillations remain relevant.

\bigskip

{\bf Acknowledgments}

It is a pleasure to thank J.W.F. Valle for useful discussions.  We
also thank K. Kainulainen for helpful comments.  The work of AD and SH
was supported in part by the Danish National Science Research Council
through grant 11-9640-1 and in part by Danmarks Grundforskningsfond
through its support of the Theoretical Astrophysical Center.  The work
of DS was supported in part by the Russian Foundation for Fundamental
Research through grants 97-02-17064A and 98-02-17493A.  SP was
supported by Conselleria d'Educaci\'o i Ci\`encia of Generalitat
Valenciana, by DGICYT grant PB95-1077 and by the EEC under the TMR
contract ERBFMRX-CT96-0090. SP and DS thank TAC for hospitality when
this work was done.

\newpage
{\large \bf Figure Captions:}
\vskip0.5cm
\noindent
{\bf Fig. 1} $~~~$ 
Asymptotic values for the effective number of neutrinos 
$N_{eff}=(\rho_{\nt}+\rho_{\nt}+\rho_{\nt}+
\rho_{\phi})/\rho_{eq}$, where $\rho_{eq}= (7\pi^2/120 ) m_0^4/x^4$ is
the energy density for one massless neutrino.  In fig.~1a we present
the region $0.1 < \tau/\mbox{sec} < 400$ and $0.1 < \mnt/\mbox{MeV} <
20$.  In fig.~1b we show the large mass region with decay times in the region
$0.01 < \tau/\mbox{sec} < 1$. In this region the effective number of
neutrinos can be less than 3.

\vskip0.5cm
\noindent
{\bf Fig. 2} $~~~$ 
For $\mnt = 0.1$ MeV and $\tnt = 1$ sec we present the evolution in time
$x$ of the energy densities (fig.~2a) and number of particles densities
(fig.~2b) in units of massless equilibrium neutrinos ($\rho_{eq}=
(7\pi^2/120 ) m_0^4/x^4$ and $n_{eq}= 0.183 m_0^3/x^3$).  In the time
region $1<x<6$ the massive $\nt$ decays to $\nu_\mu$ and the majoron $\phi$,
but because of the small mass, the  inverse decay-process delays the
complete vanishing of $\nt$ until time
$x=25$.

\vskip0.5cm
\noindent
{\bf Fig. 3} $~~~$ 
For $\mnt = 10$ MeV and $\tnt = 10$ sec we present the evolution in time
$x$ of energy densities (fig.~3a) and number of particles densities
(fig.~3b) in units of massless equilibrium neutrinos ($\rho_{eq}=
(7\pi^2/120 ) m_0^4/x^4$ and $n_{eq}=0.183 m_0^3/x^3$).  In the time
region $0.1<x<3$ the decay does not play any role, in the region $x<1$
the annihilation of $\nt$ into other particles is important.  
Massive $\nt$ decays to $\nu_\mu$ and majoron $\phi$ around $x=3$.

\vskip0.5cm
\noindent
{\bf Fig. 4} $~~~$ 
Relative number of equivalent massless neutrino species, $\Delta N =
N_{eqv} - 3$, as a function of $\nt$ mass and  lifetime $\tau$, found
from $^4$He. In fig.~4a we present a 2-dimensional surface for $\Delta
N(\mnt,\tau)$. In fig.~4b $\Delta N$ as a function of $\nt$ mass for
different lifetimes including the asymptotic value for stable massive
$\nt$ (decay time $\tau=\infty$) is presented.

\vskip0.5cm
\noindent
{\bf Fig. 5} $~~~$
$^4$He for different lifetimes as a function of the $\nt$ mass. 
$\eta_{10} = 3$ (fig.~5a). $^2$H for different lifetimes  (fig.~5b).
$^7$Li for different lifetimes (fig.~5c).

\vskip0.5cm
\noindent
{\bf Fig. 6} $~~~$
Fig.~6a shows $\Delta N$ contours found from $^4$He as a function of
$\nt$ mass and lifetime.  $\eta_{10} = 3$. 
The contours
shown correspond to  $\Delta N =$
-0.6, -0.4, -0.2, -0.1,  0.0, 0.1, 0.2, 0.4, 0.6.
Fig.~6b shows the same 
as fig.~6a with small lifetimes $0.001 \leq \tau/\mbox{sec} \leq 0.1$.

\vskip1.5cm
\begin{center}$\nt \rightarrow \nu_e + \phi$
\end{center}

\vskip0.5cm
\noindent
{\bf Fig. 7a} $~~~$
$\Delta N$  surface  found from $^4$He as  a function of
$\nt$ mass and lifetime. $\eta_{10} = 3$.

\vskip0.5cm
\noindent
{\bf Fig. 7b} $~~~$
$\Delta N$ for different lifetimes as  a function of
$\nt$ mass. $\eta_{10} = 3$.

\vskip0.5cm
\noindent
{\bf Fig. 8a} $~~~$
$\Delta N$ contours for  $\eta_{10} = 3$ as a function of
$\nt$ mass and lifetime. Contours follow the  increase with
the step of $0.2$.

\vskip0.5cm
\noindent
{\bf Fig. 8b} $~~~$
Same as fig.~8a with small lifetimes $0.001 \leq \tau/\mbox{sec} \leq 0.1$.

\vskip0.5cm
\noindent
{\bf Fig. 8c} $~~~$
$^4$He contours for  $\eta_{10} = 3$ as a function of
$\nt$ mass and lifetime. Contours follow the increase with
the step of $0.05$.
 
\vskip0.5cm
\noindent
{\bf Fig. 9} $~~~$
$^4$He (9a), $^2$H (9b) and $^7$Li (9c) as functions of the $\nt$
mass for different lifetimes. $\eta_{10} = 3$.

\vskip0.5cm
\noindent
{\bf Fig. 10} $~~~$
$^4$He (10a), $^2$H (10b) and $^7$Li (10c) as functions of the $\nt$
mass for different lifetimes. $\eta_{10} = 5$.

\vskip0.5cm
\noindent
{\bf Fig. 11} $~~~$
$^4$He (11a), $^2$H (11b) and $^7$Li (11c) as functions of the $\nt$
mass for different lifetimes. $\eta_{10} = 7$.

\vskip0.5cm   
\noindent
{\bf Fig. 12 } $~~~$
Energy density of $\nt$ and $\phi$ as functions of time $x$ 
for the cases with 
$ f_\phi (x_{in}) = 1/(\exp(y)-1)$ and $ f_\phi (x_{in}) =0$.

\newpage
\psfig{file=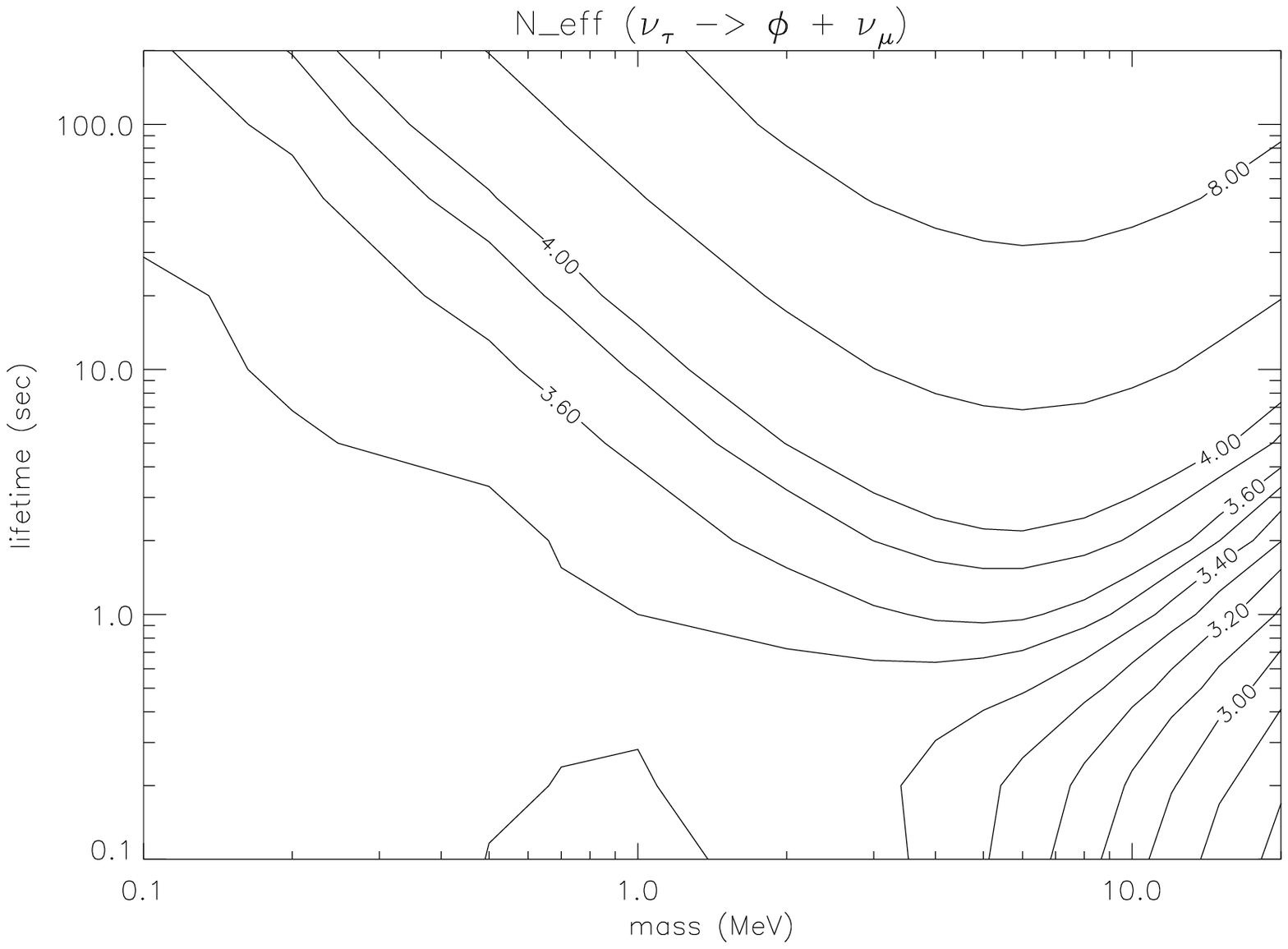,width=5in,height=3.5in}
\begin{center}
{\bf Figure 1a.}
\end{center}

\psfig{file=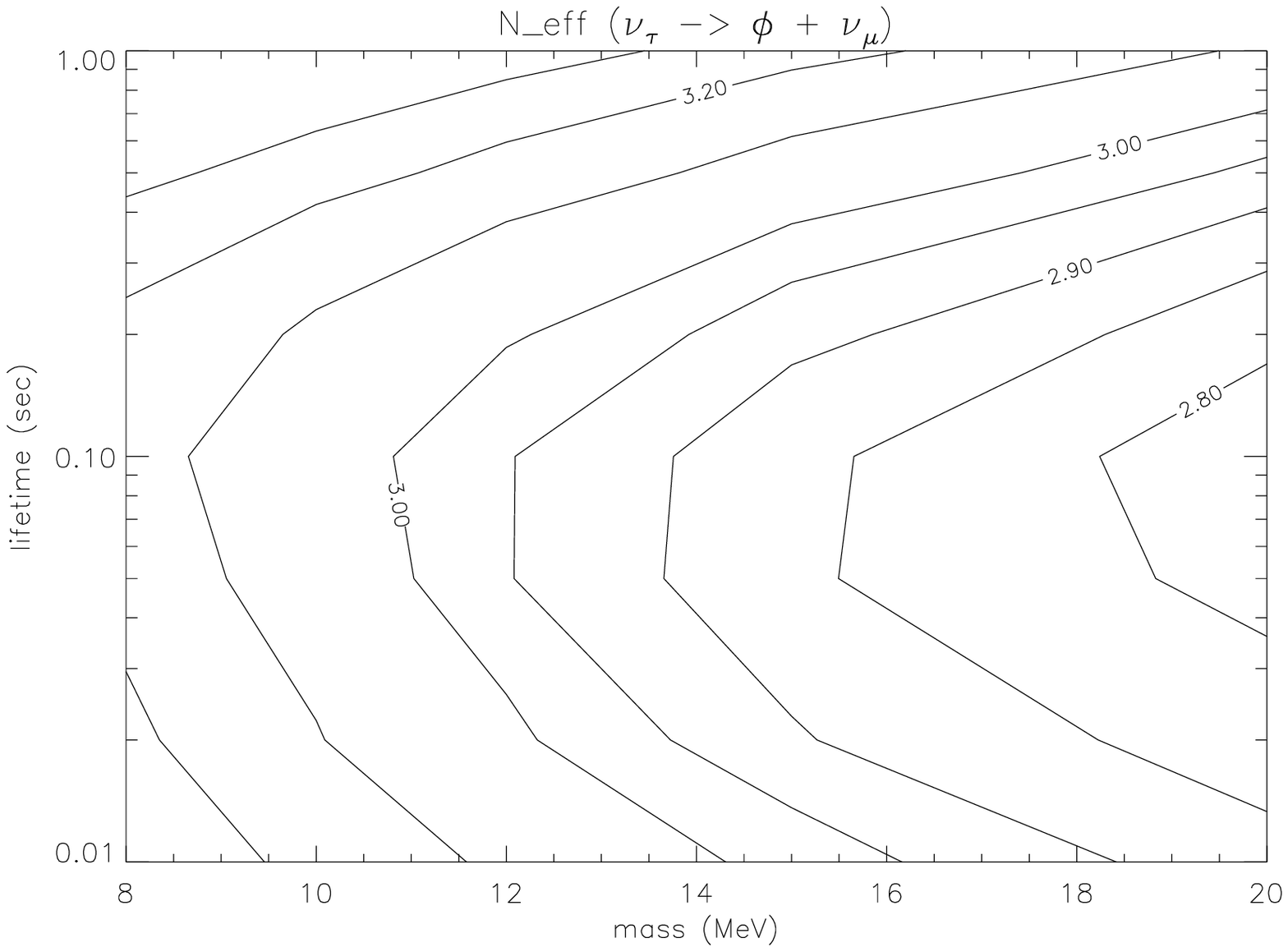,width=5in,height=3.5in}
\begin{center}
{\bf Figure 1b.}
\end{center}

\newpage
\psfig{file=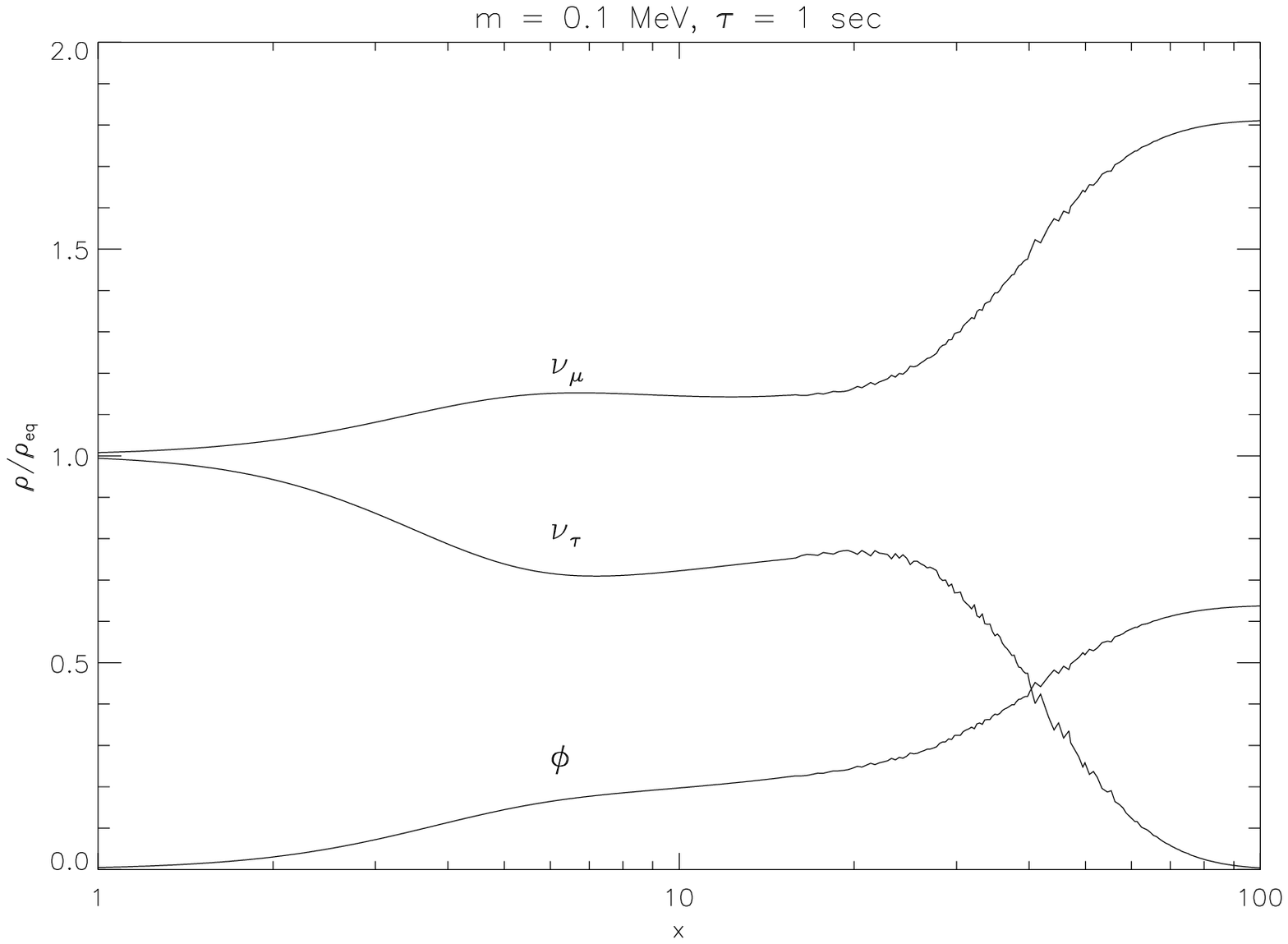,width=5in,height=3.5in}
\begin{center}
{\bf Figure 2a.}
\end{center}

\psfig{file=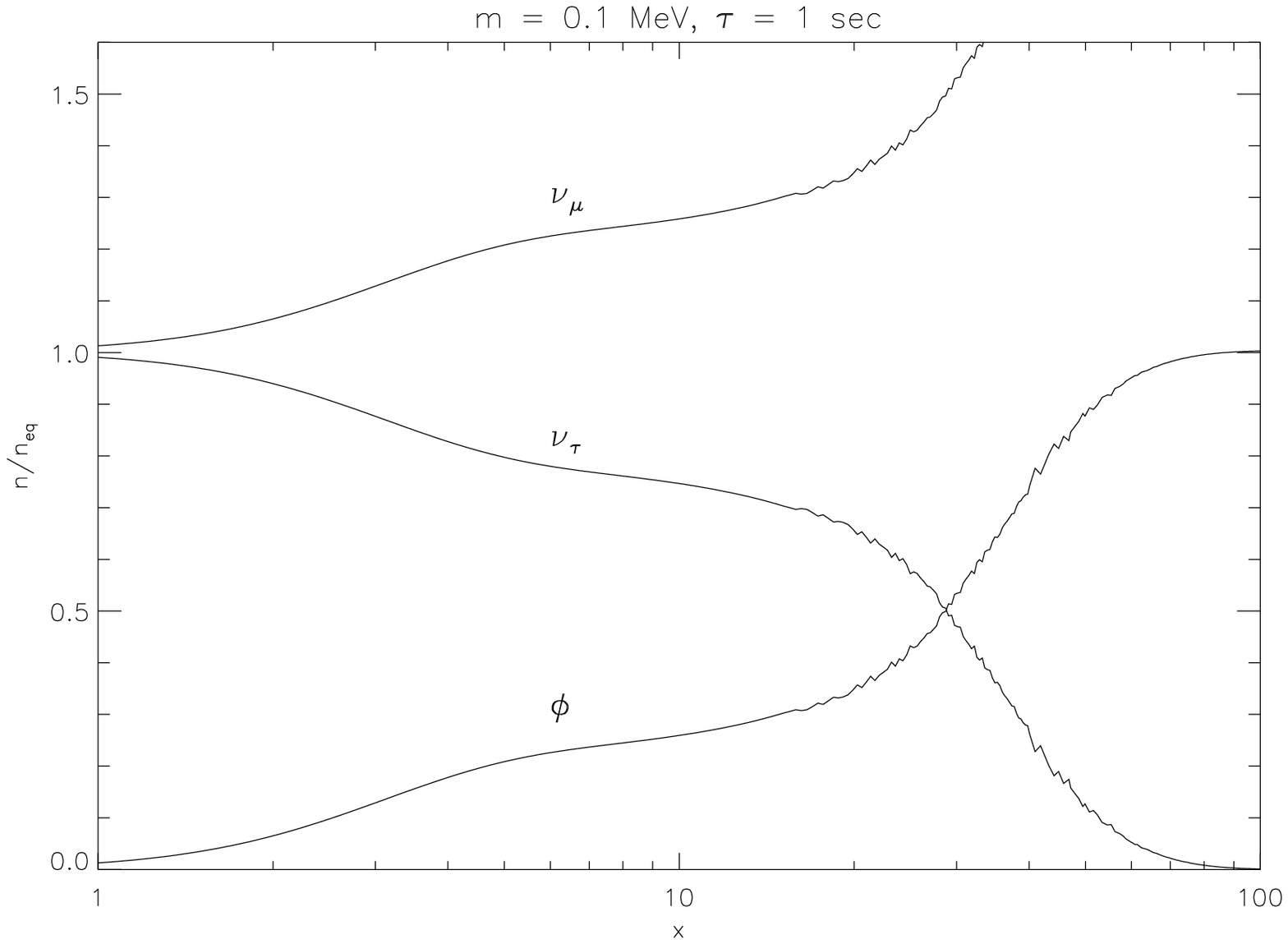,width=5in,height=3.5in}
\begin{center}
{\bf Figure 2b.}
\end{center}

\newpage
\psfig{file=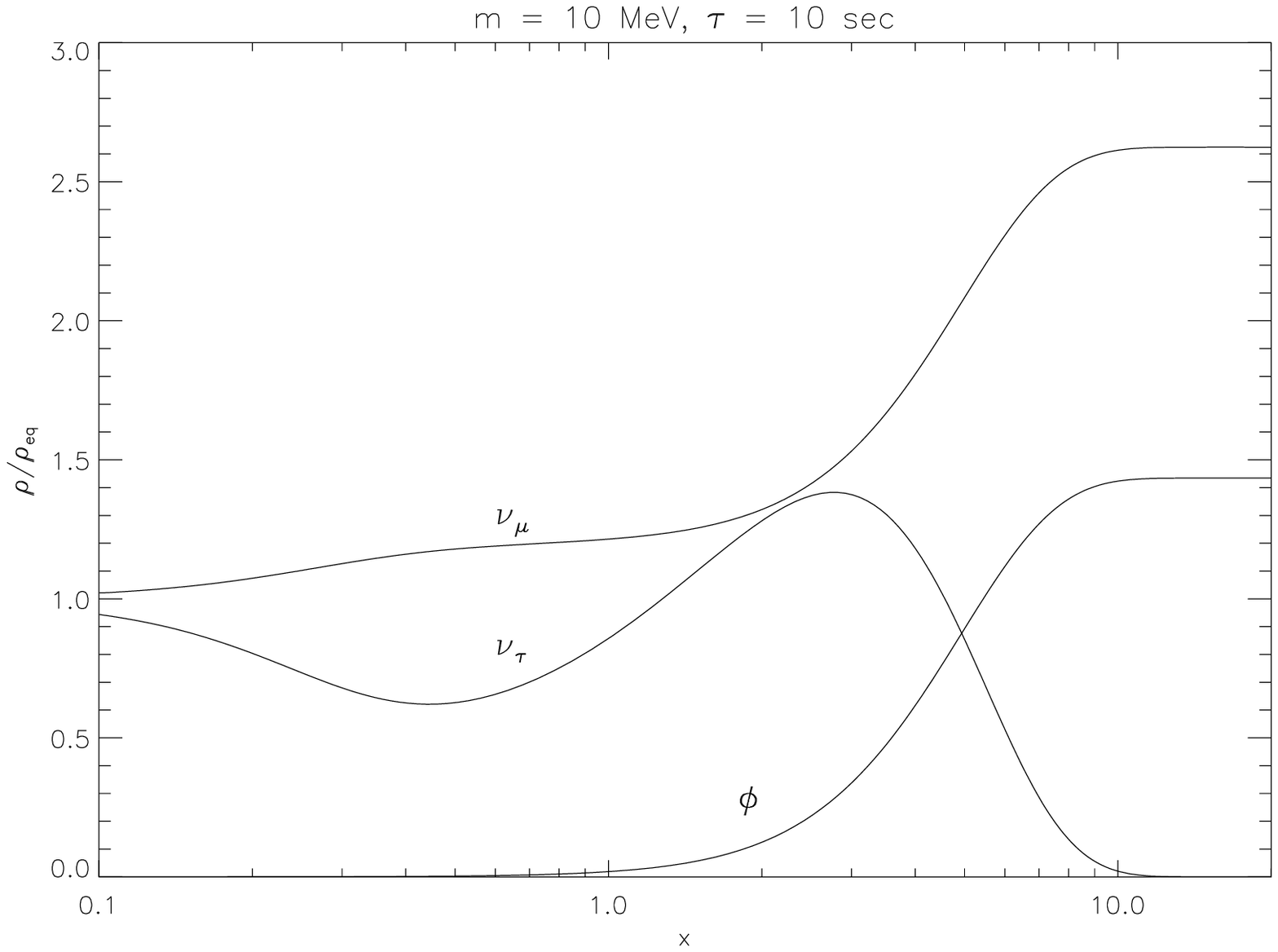,width=5in,height=3.5in}
\begin{center}
{\bf Figure 3a.}
\end{center}

\psfig{file=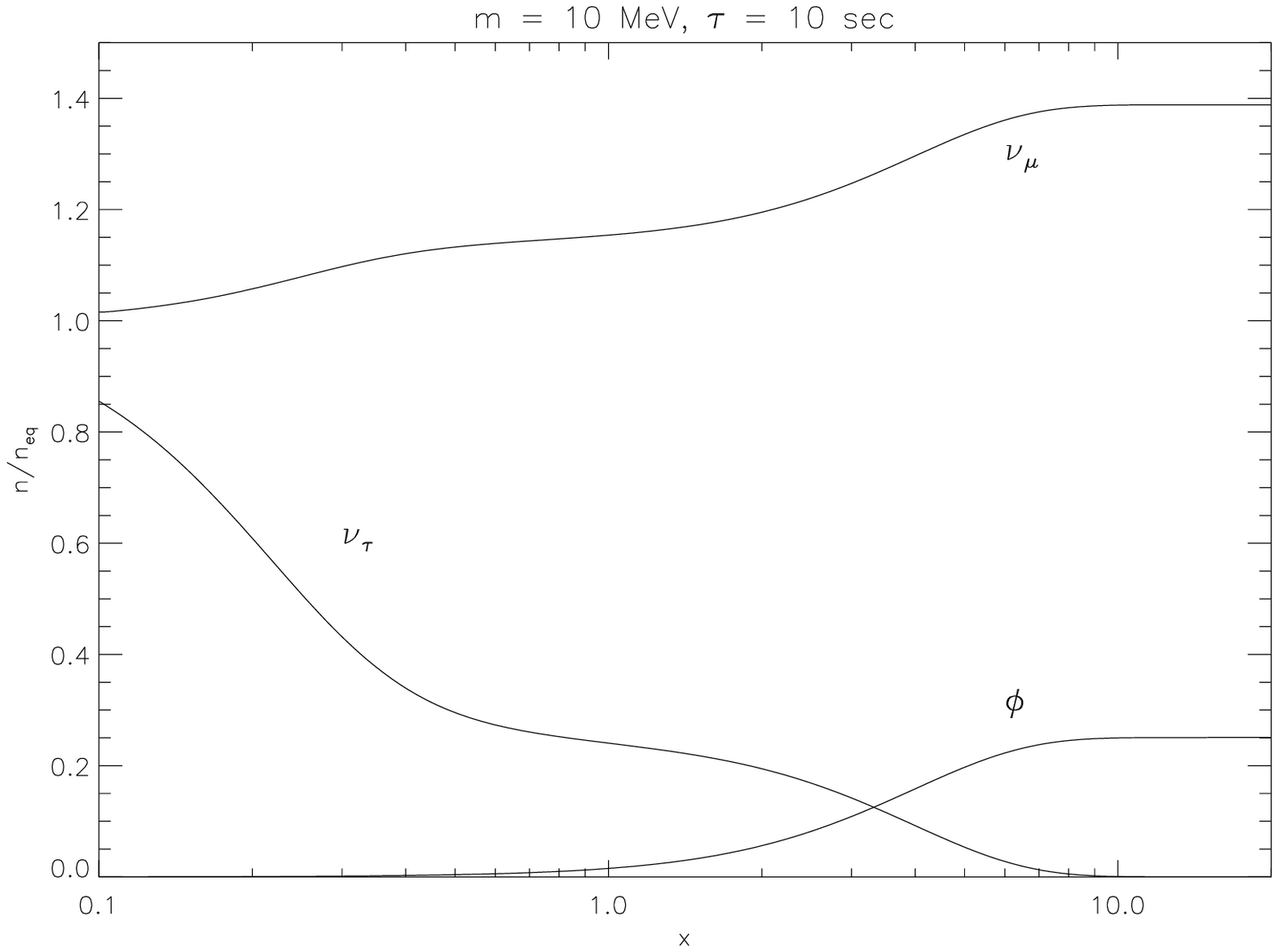,width=5in,height=3.5in}
\begin{center}
{\bf Figure 3b.}
\end{center}

\newpage

\psfig{file=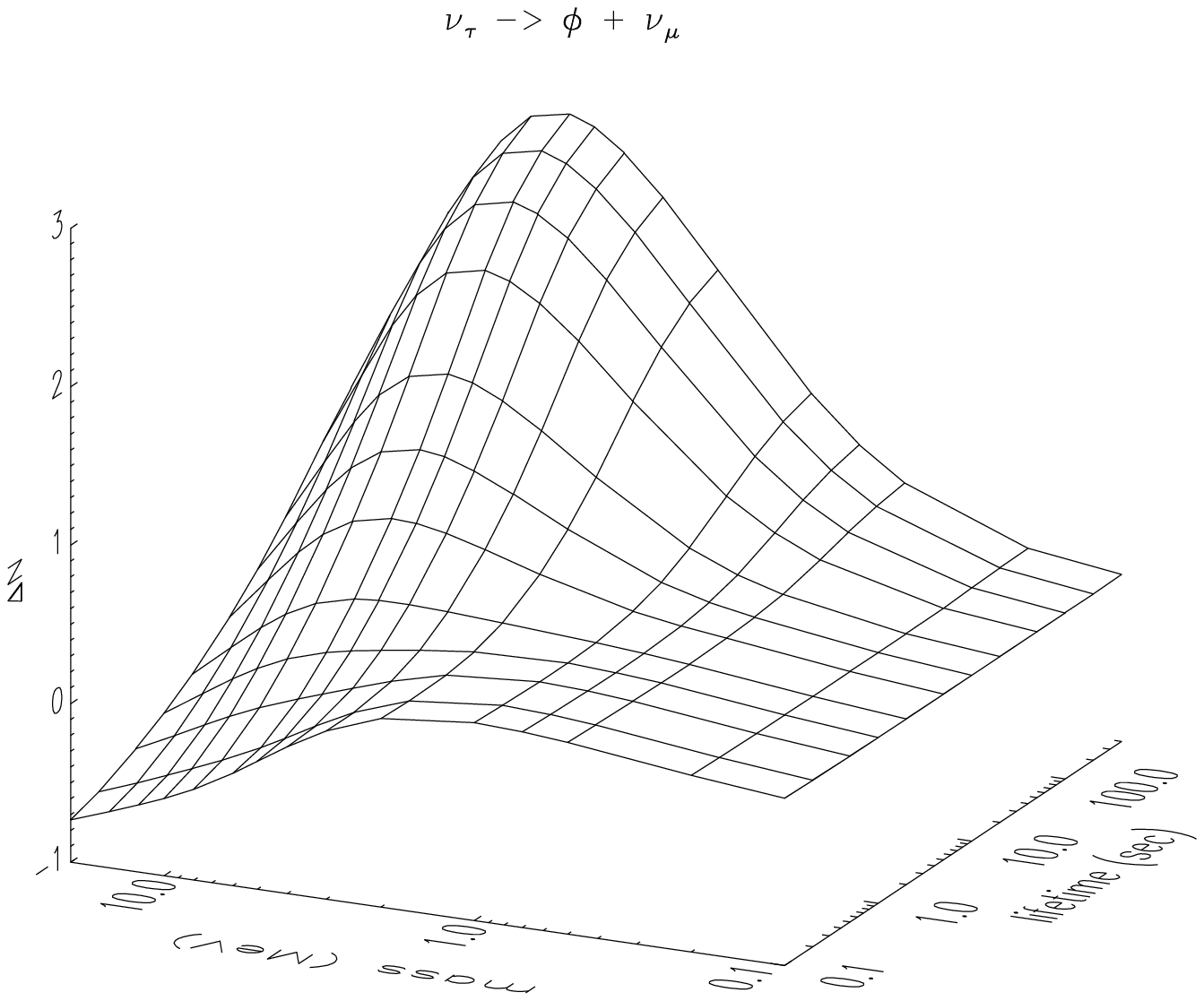,width=5in,height=3.5in}
\begin{center}
{\bf Figure 4a.}
\end{center}

\psfig{file=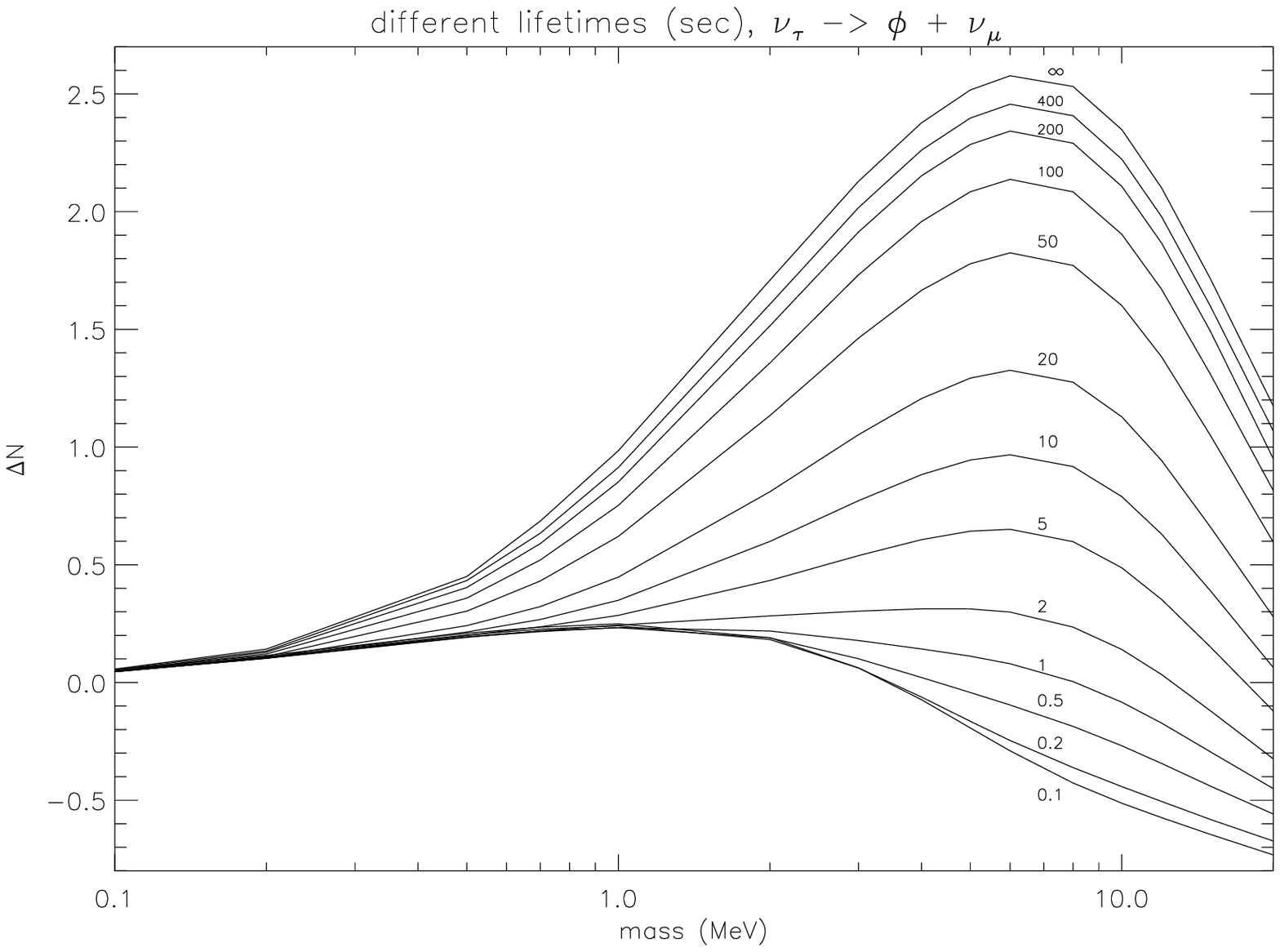,width=5in,height=3.5in}
\begin{center}
{\bf Figure 4b.}
\end{center}

\newpage

\psfig{file=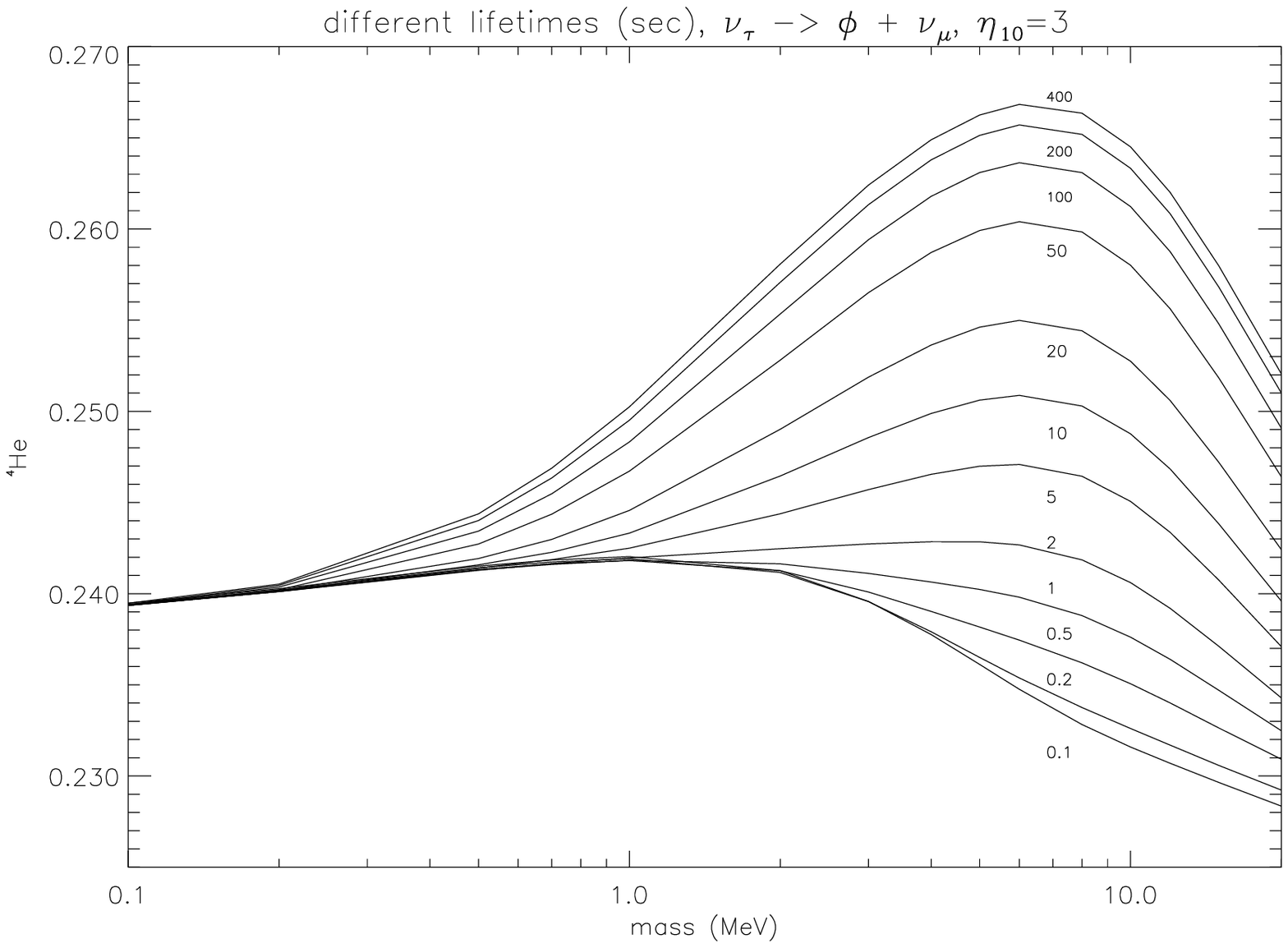,width=5in,height=3.5in}
\begin{center}
{\bf Figure 5a.}
\end{center}

\psfig{file=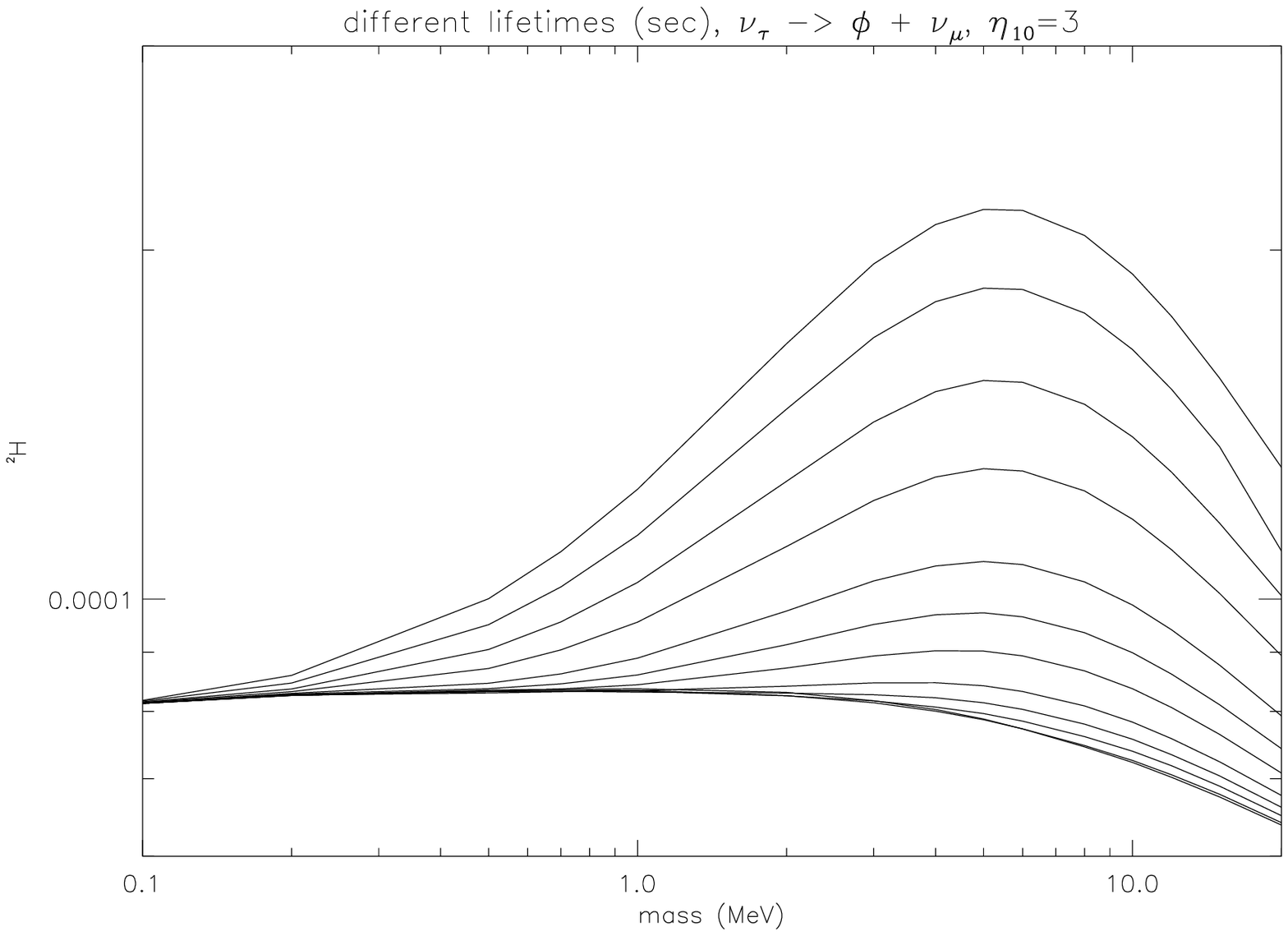,width=5in,height=3.5in}
\begin{center}
{\bf Figure 5b.}
\end{center}

\newpage

\psfig{file=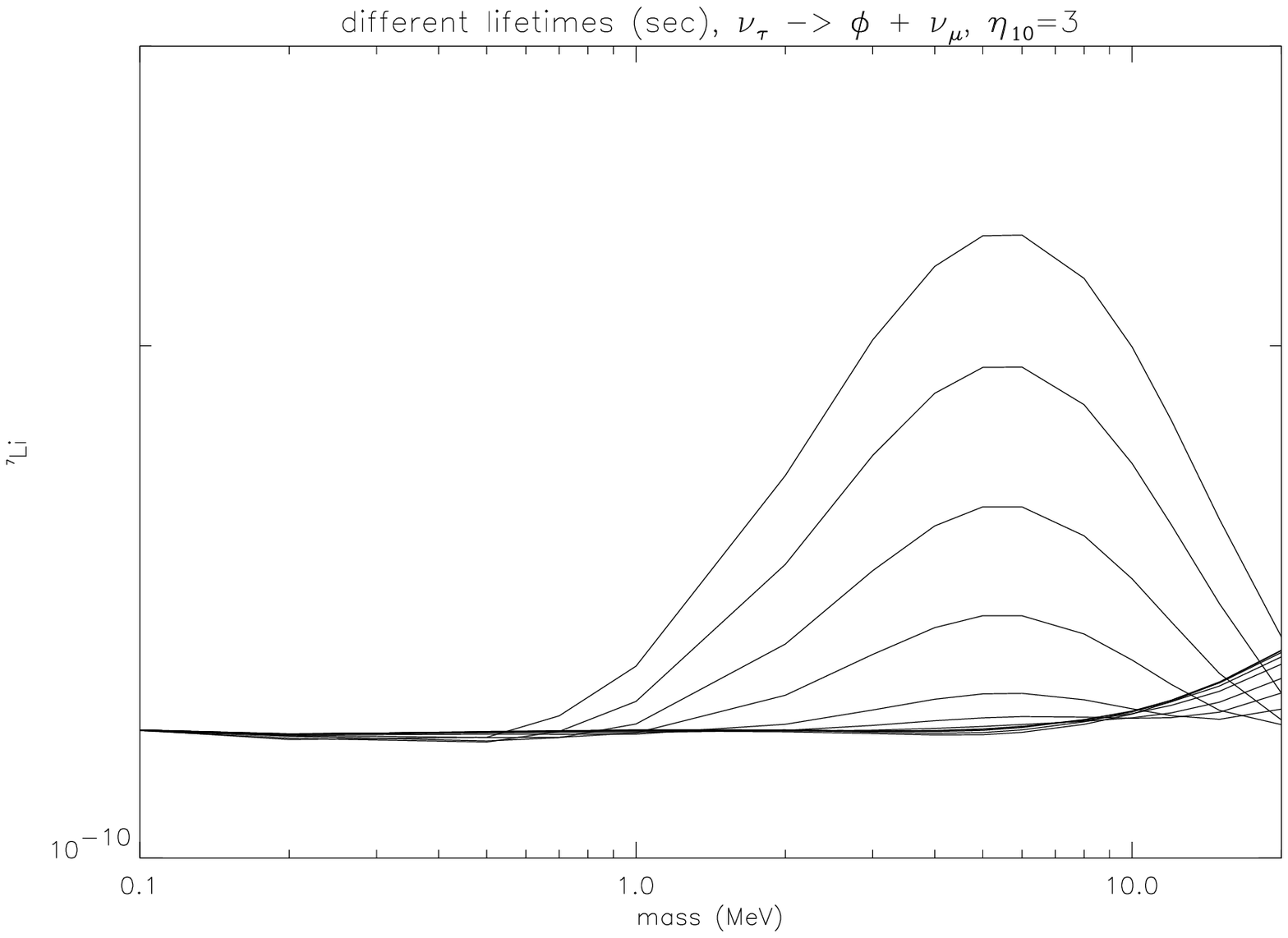,width=5in,height=3.5in}
\begin{center}
{\bf Figure 5c.}
\end{center}

\psfig{file=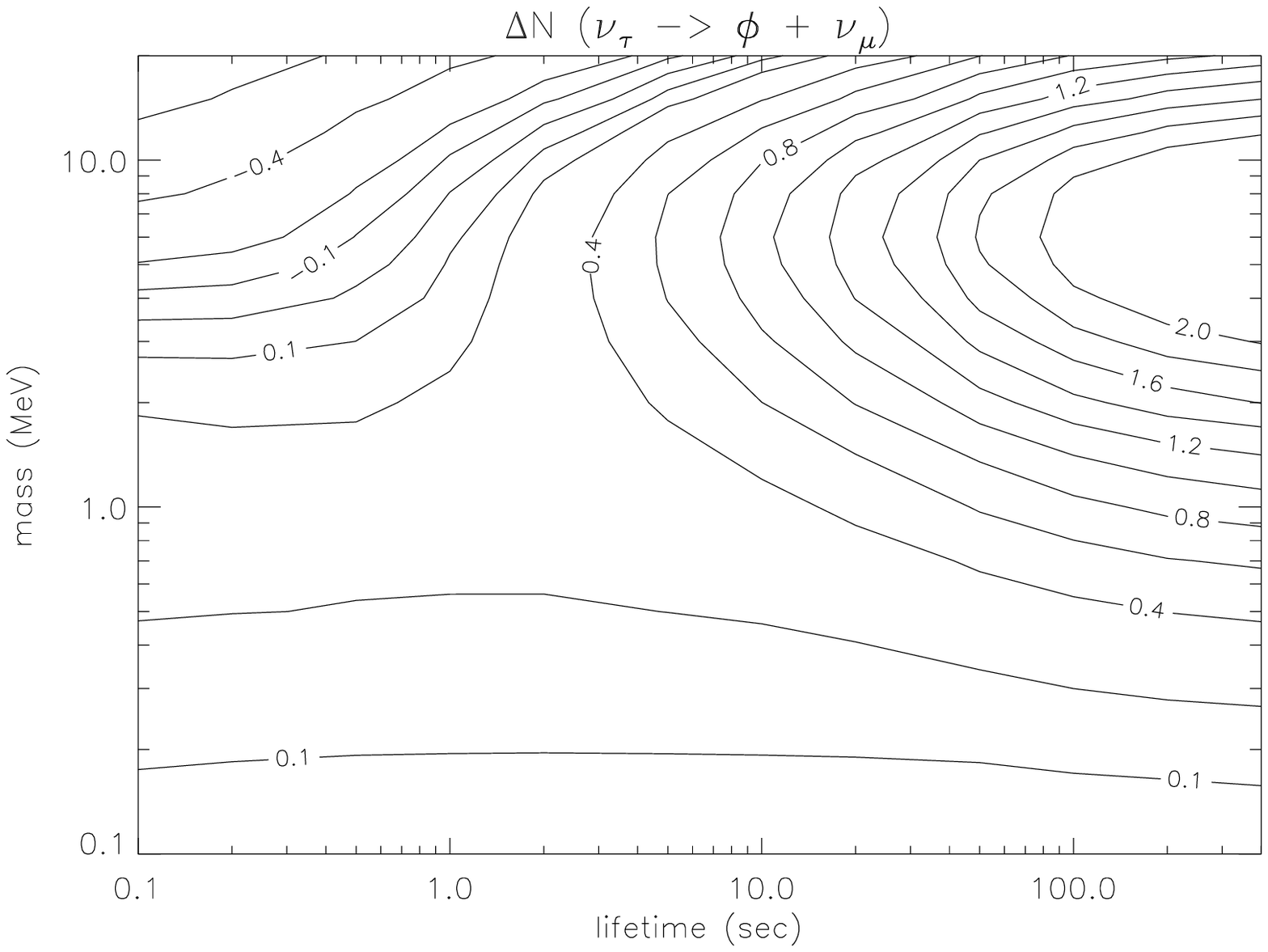,width=5in,height=3.5in}
\begin{center}
{\bf Figure 6a.}
\end{center}

\newpage

\psfig{file=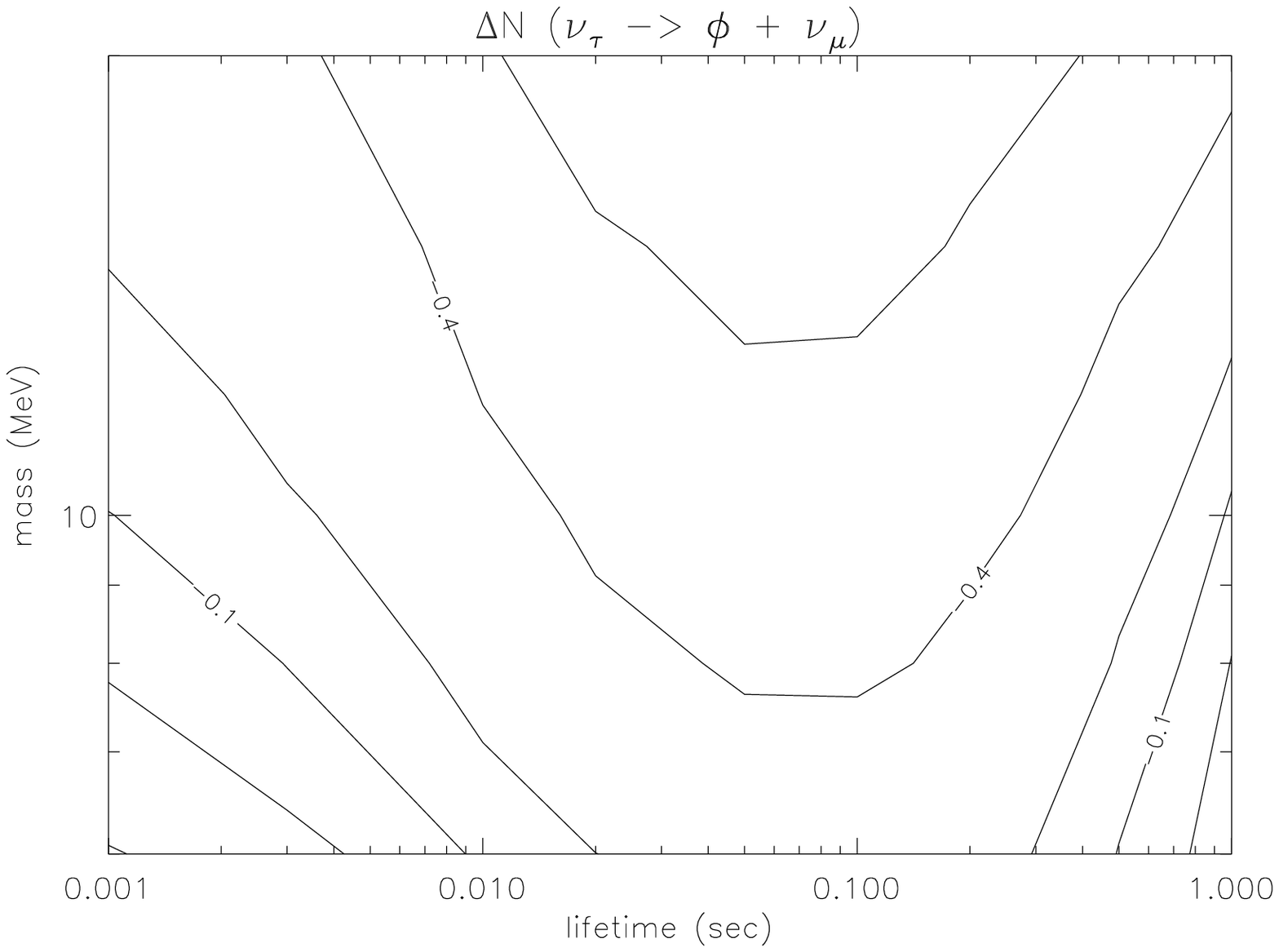,width=5in,height=3.5in}
\begin{center}
{\bf Figure 6b.}
\end{center}

\psfig{file=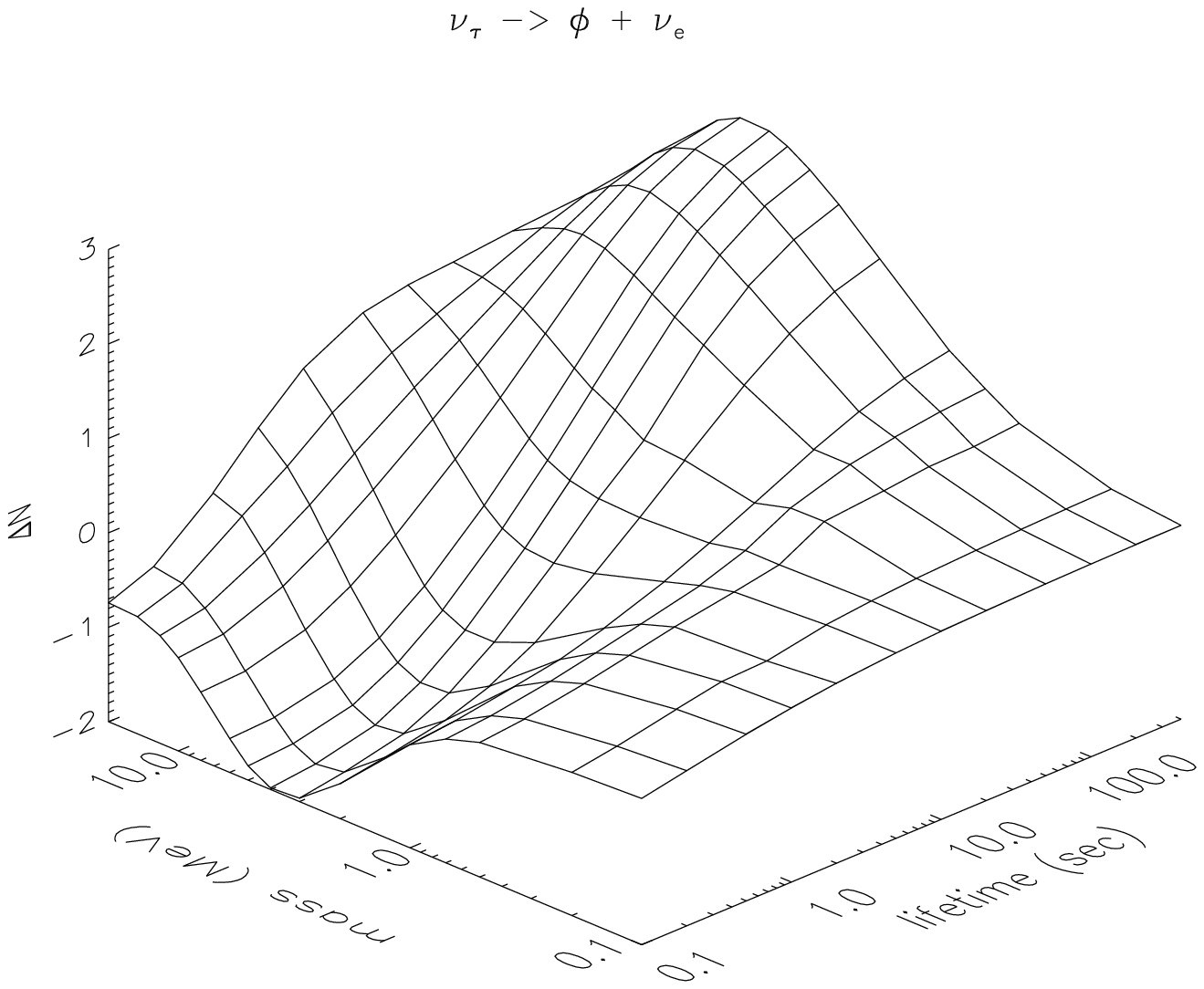,width=5in,height=3.5in}
\begin{center}
{\bf Figure 7a.}
\end{center}

\newpage

\psfig{file=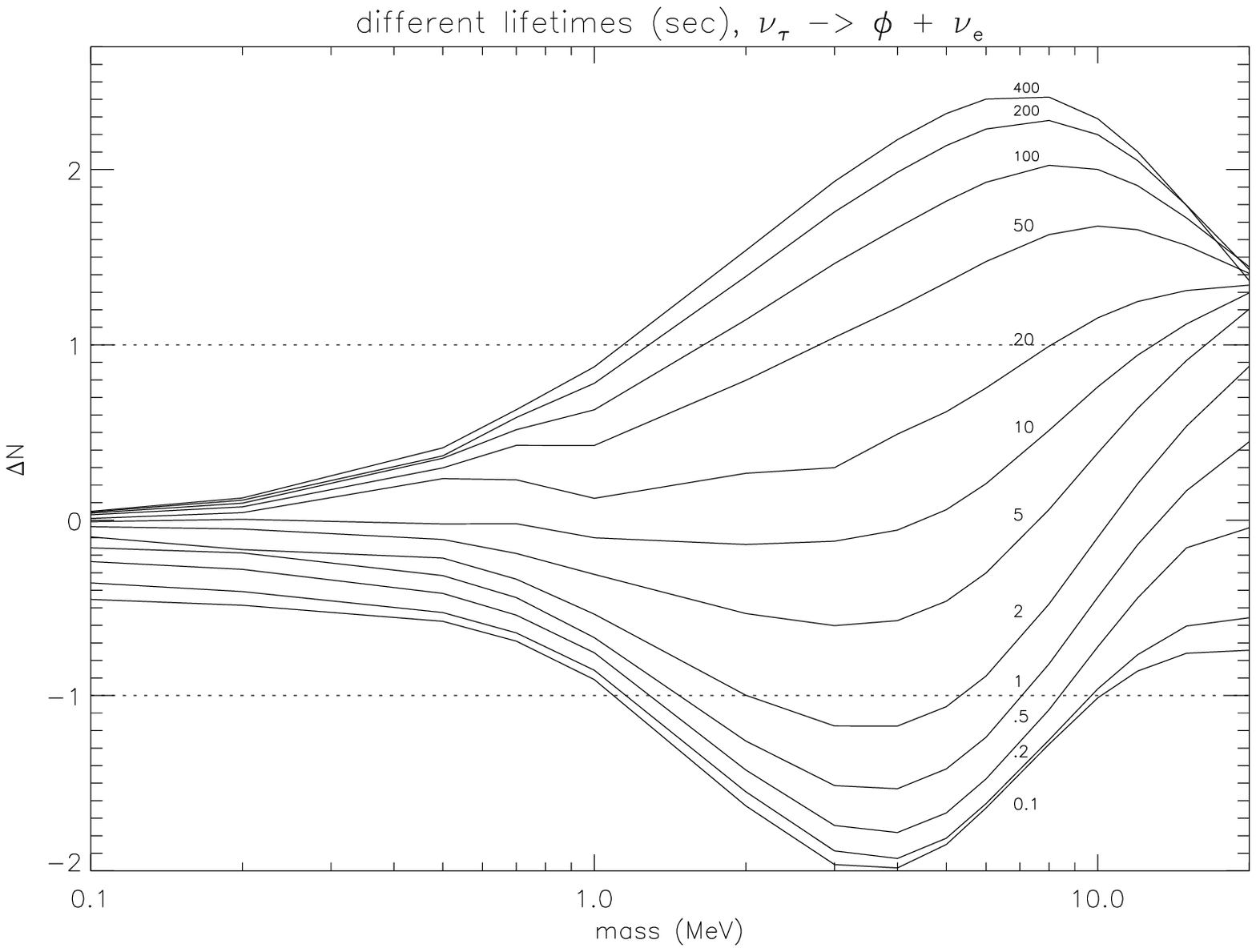,width=5in,height=3.5in}
\begin{center}
{\bf Figure 7b.}
\end{center}

\psfig{file=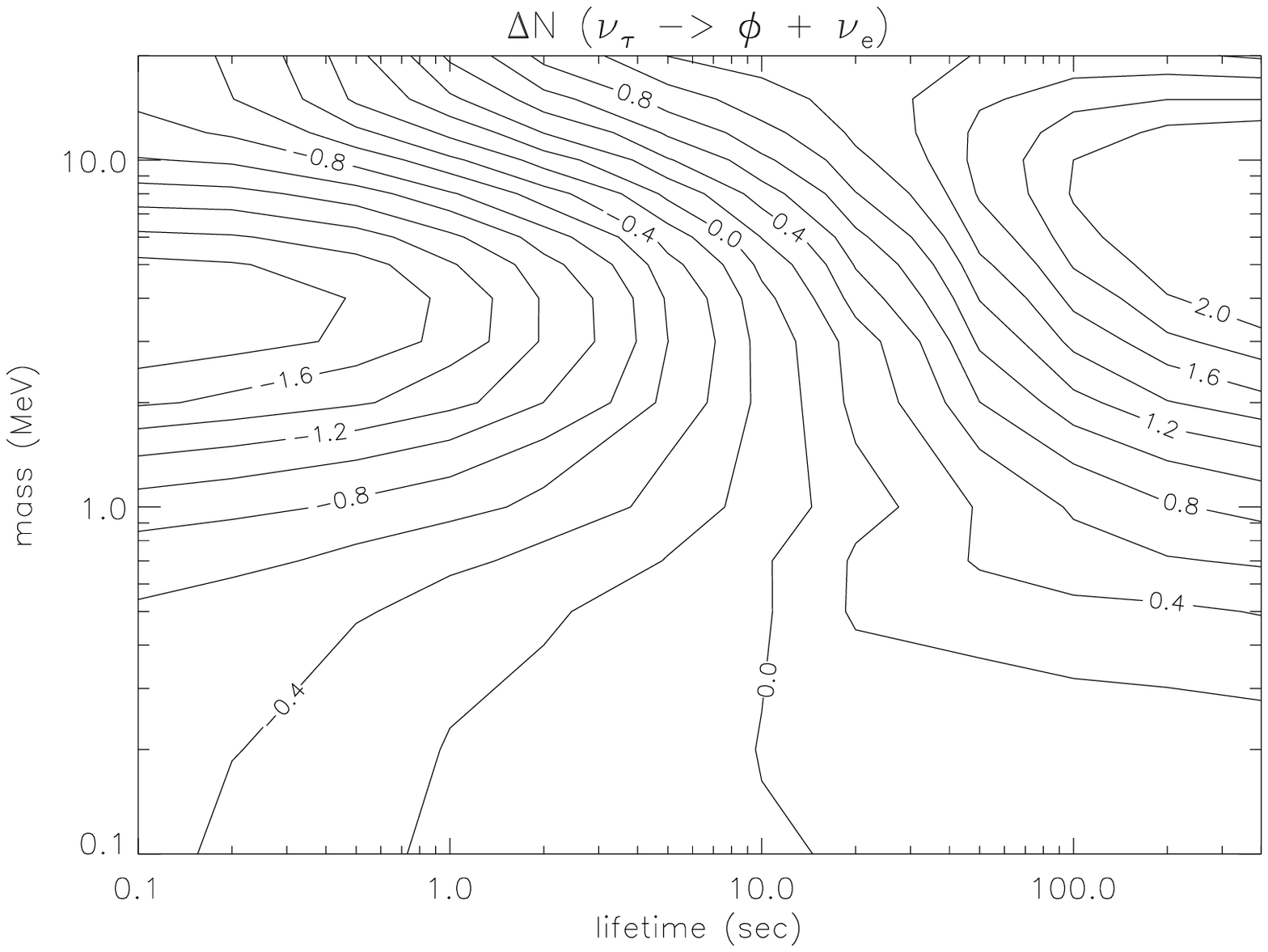,width=5in,height=3.5in}
\begin{center}
{\bf Figure 8a.}
\end{center}

\newpage

\psfig{file=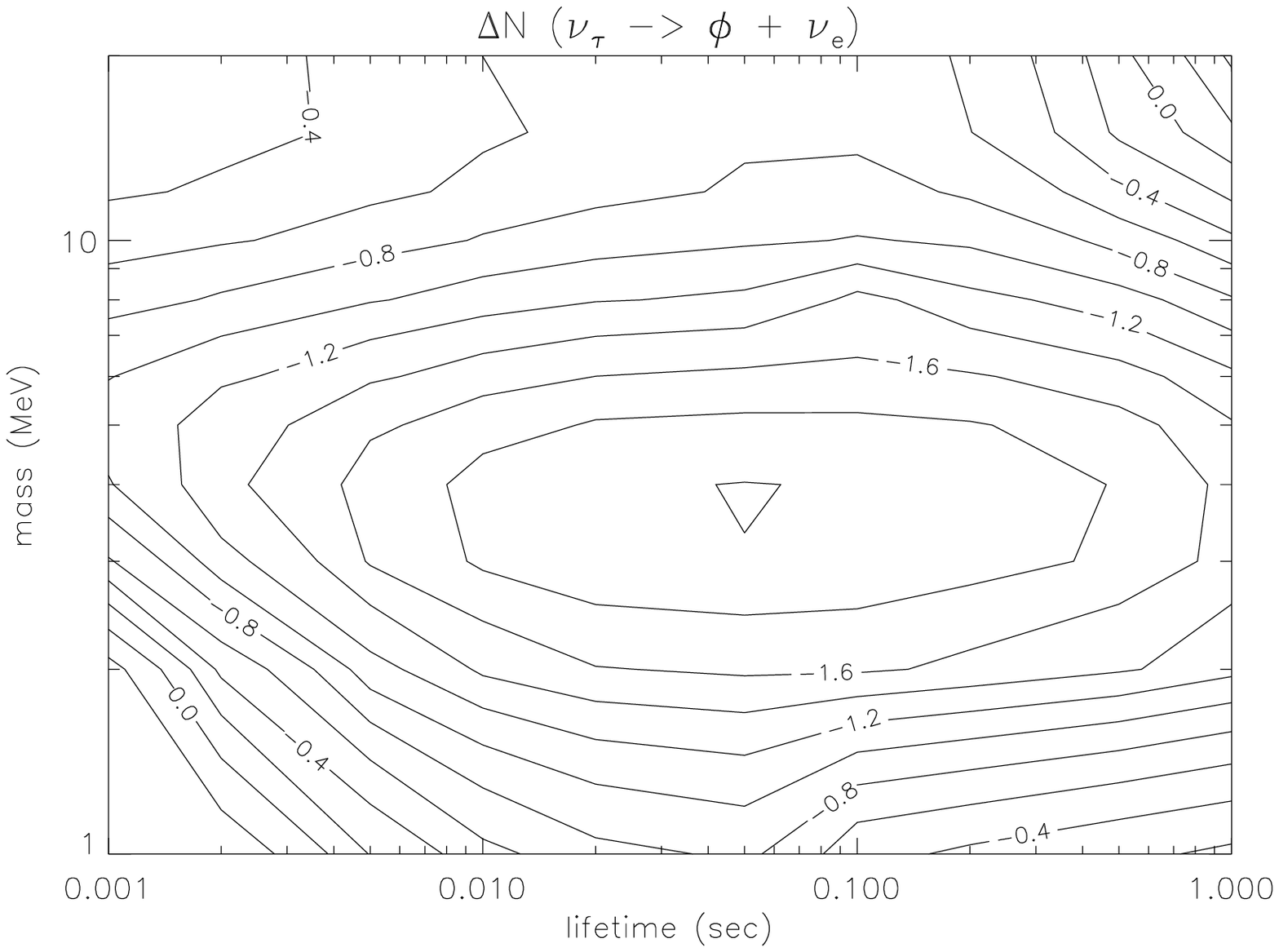,width=5in,height=3.5in}
\begin{center}
{\bf Figure 8b.}
\end{center}

\psfig{file=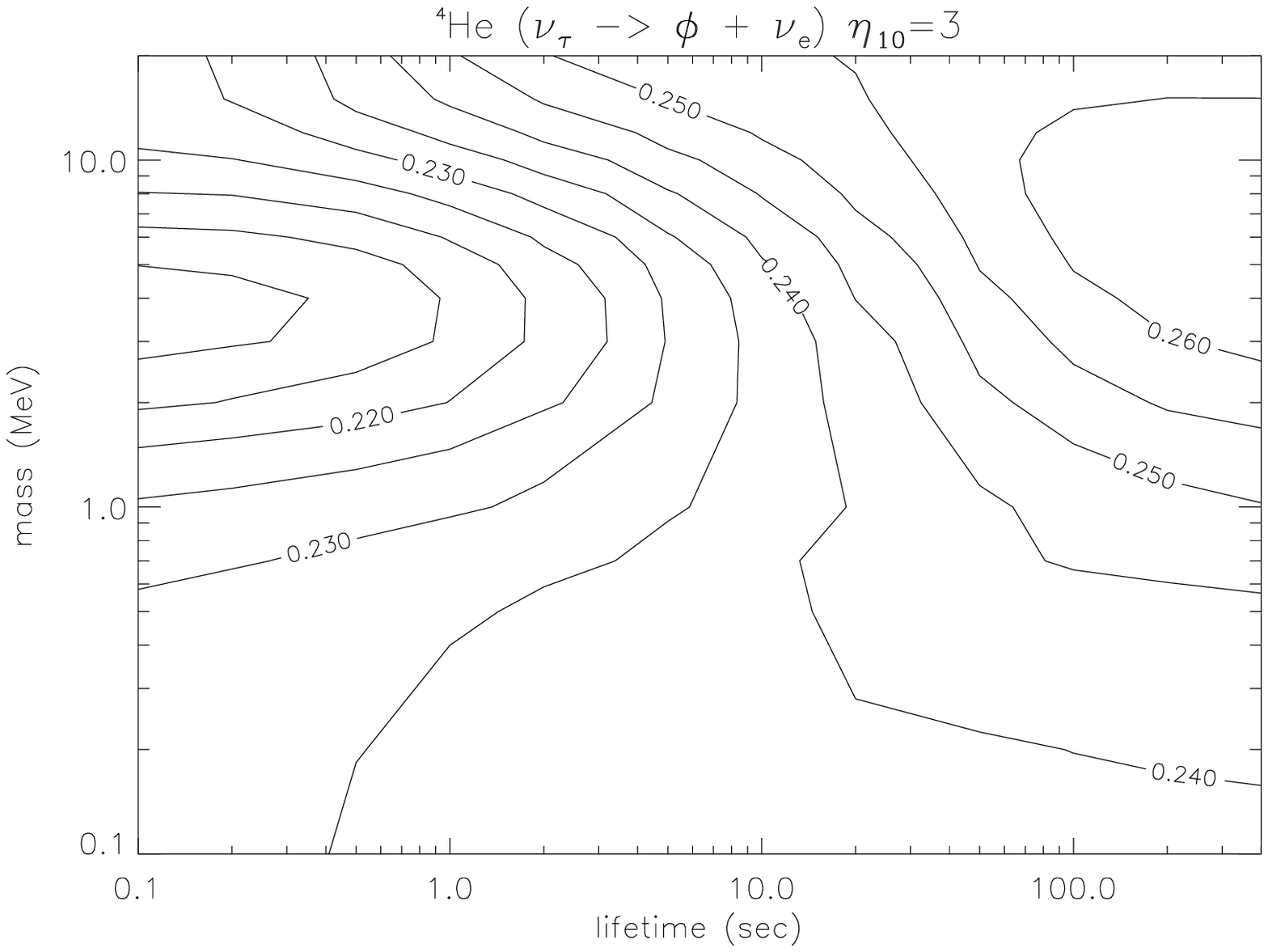,width=5in,height=3.5in}
\begin{center}
{\bf Figure 8c.}
\end{center}

\newpage

\psfig{file=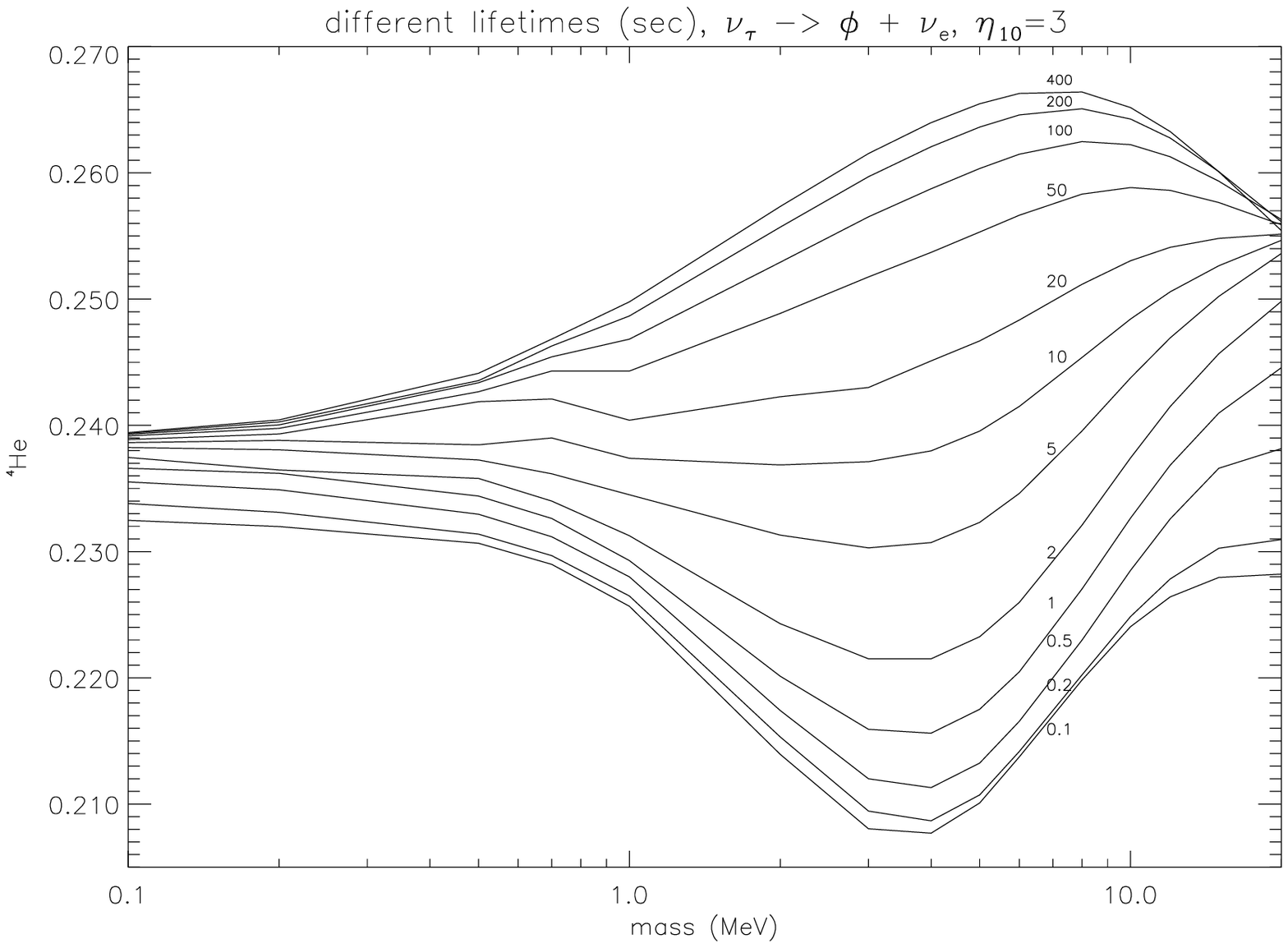,width=5in,height=3.5in}
\begin{center}
{\bf Figure 9a.}
\end{center}

\psfig{file=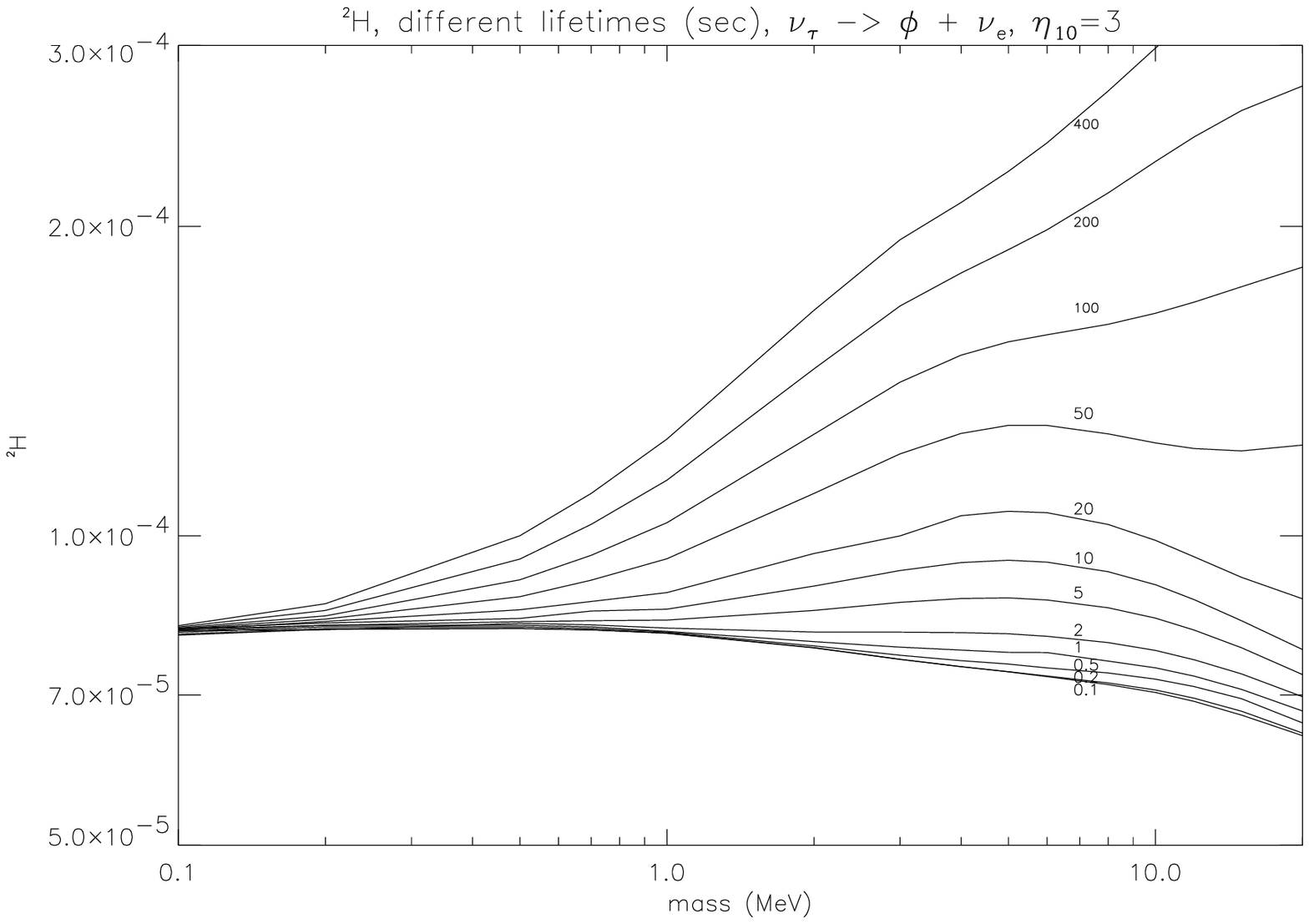,width=5in,height=3.5in}
\begin{center}
{\bf Figure 9b.}
\end{center}

\newpage

\psfig{file=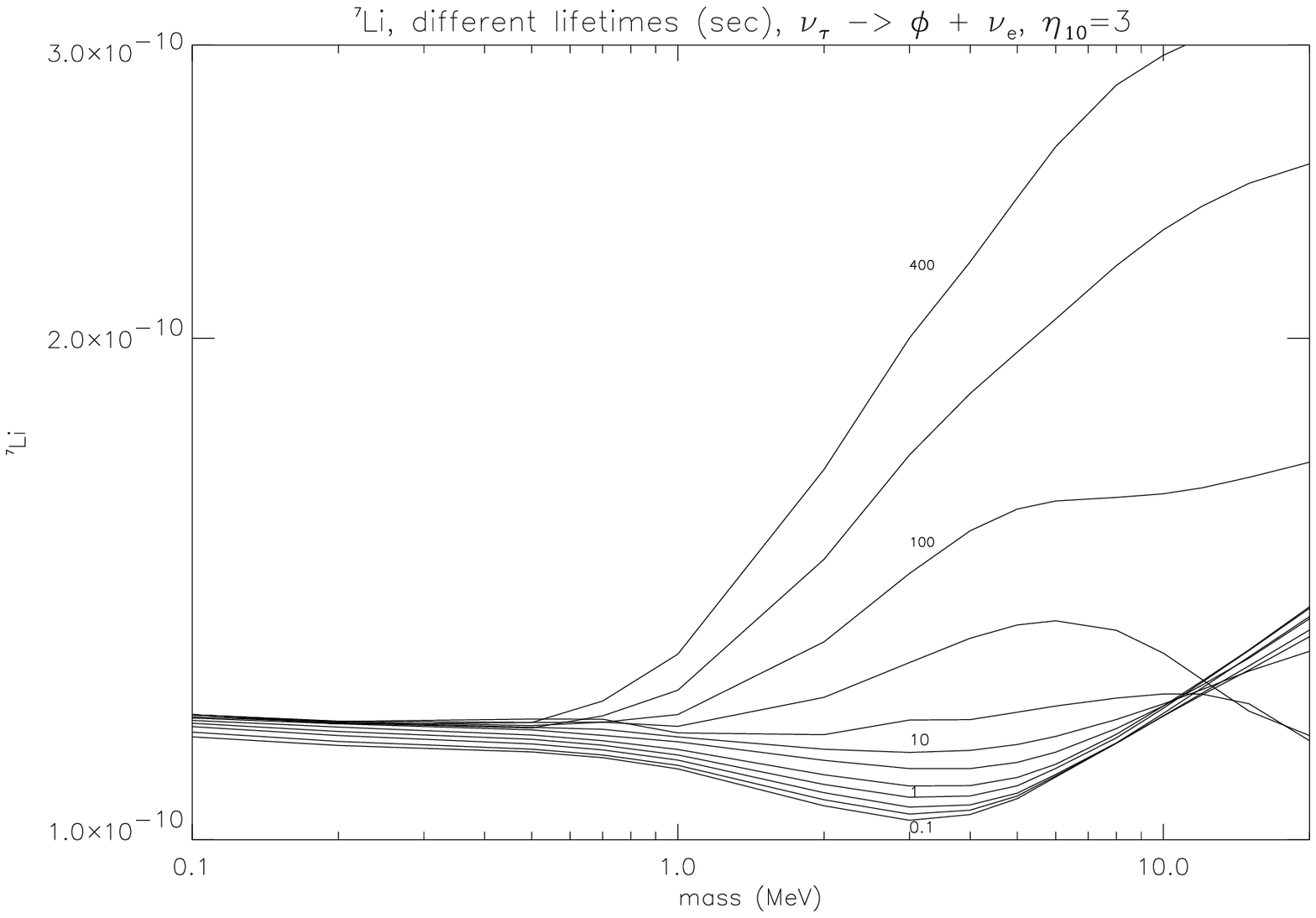,width=5in,height=3.5in}
\begin{center}
{\bf Figure 9c.}
\end{center}

\psfig{file=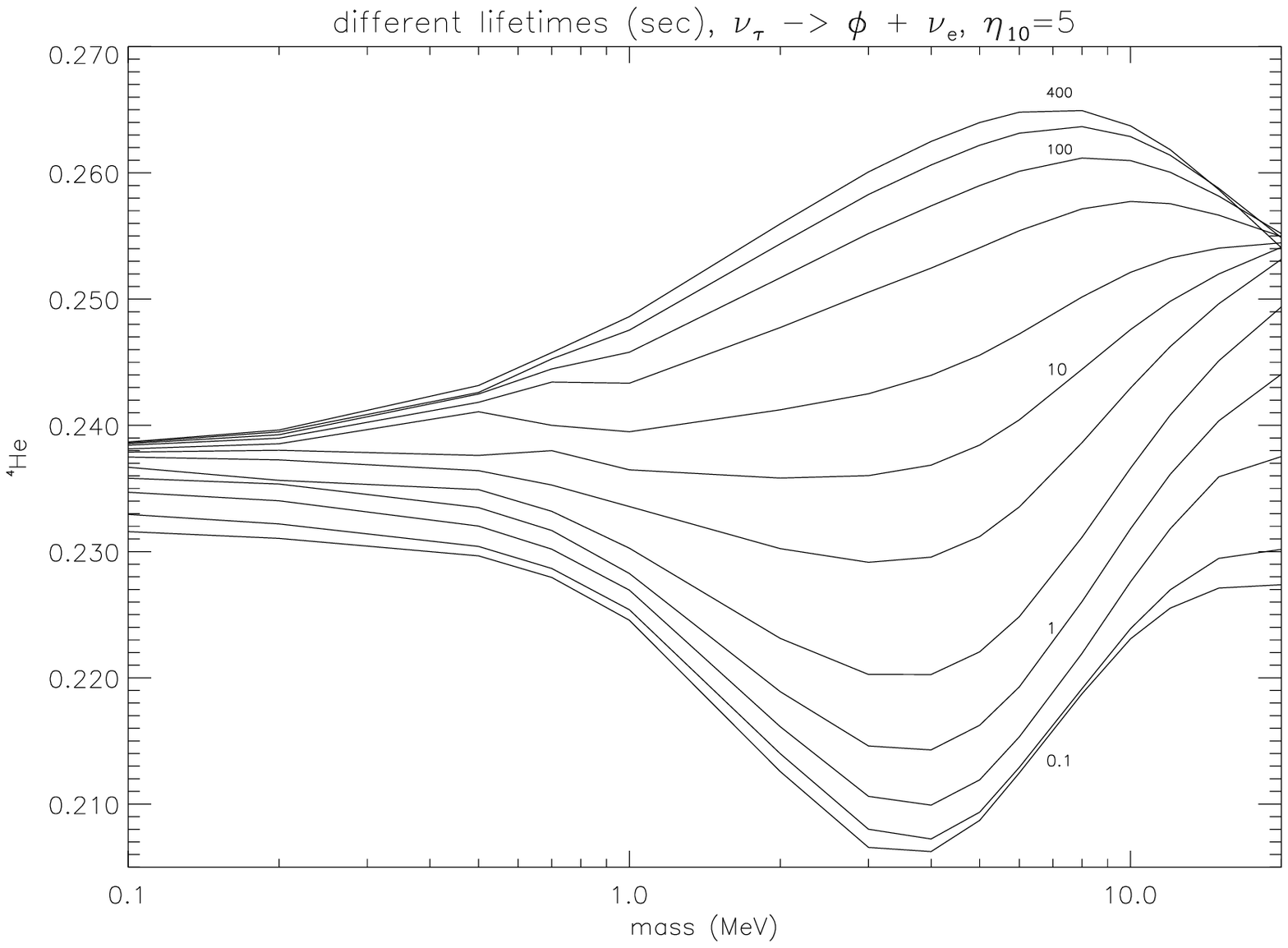,width=5in,height=3.5in}
\begin{center}
{\bf Figure 10a.}
\end{center}

\newpage

\psfig{file=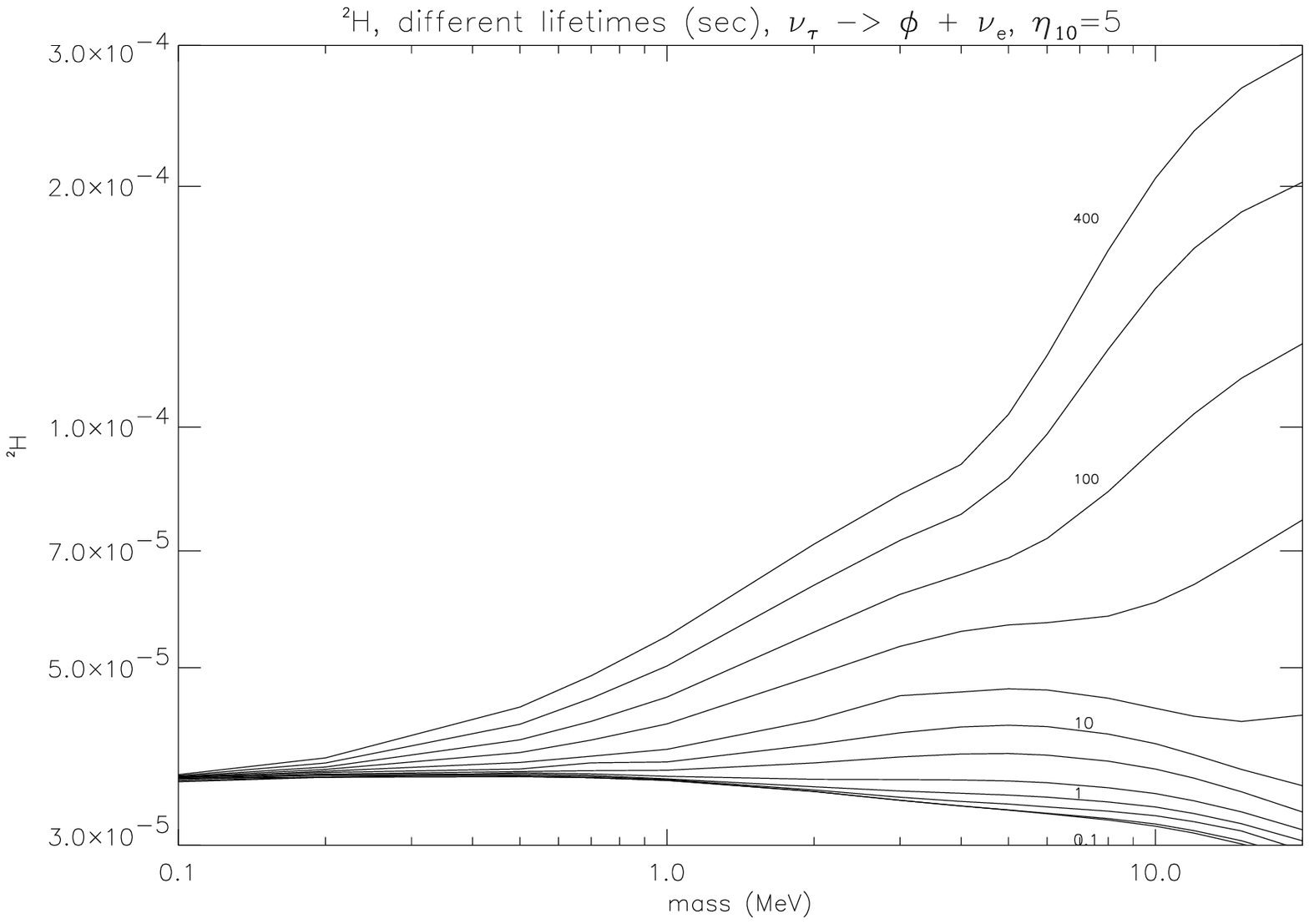,width=5in,height=3.5in}
\begin{center}
{\bf Figure 10b.}
\end{center}

\psfig{file=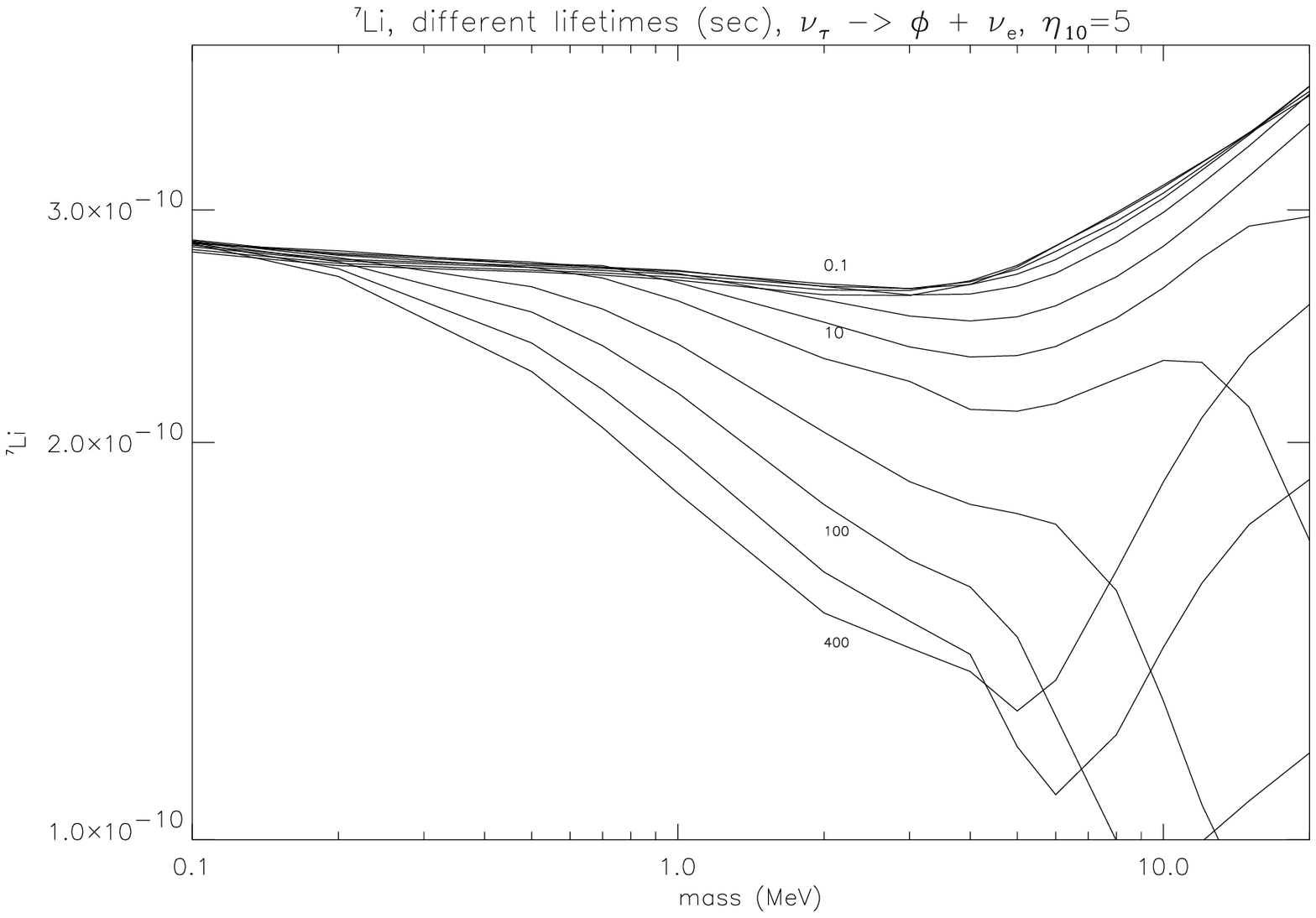,width=5in,height=3.5in}
\begin{center}
{\bf Figure 10c.}
\end{center}

\newpage

\psfig{file=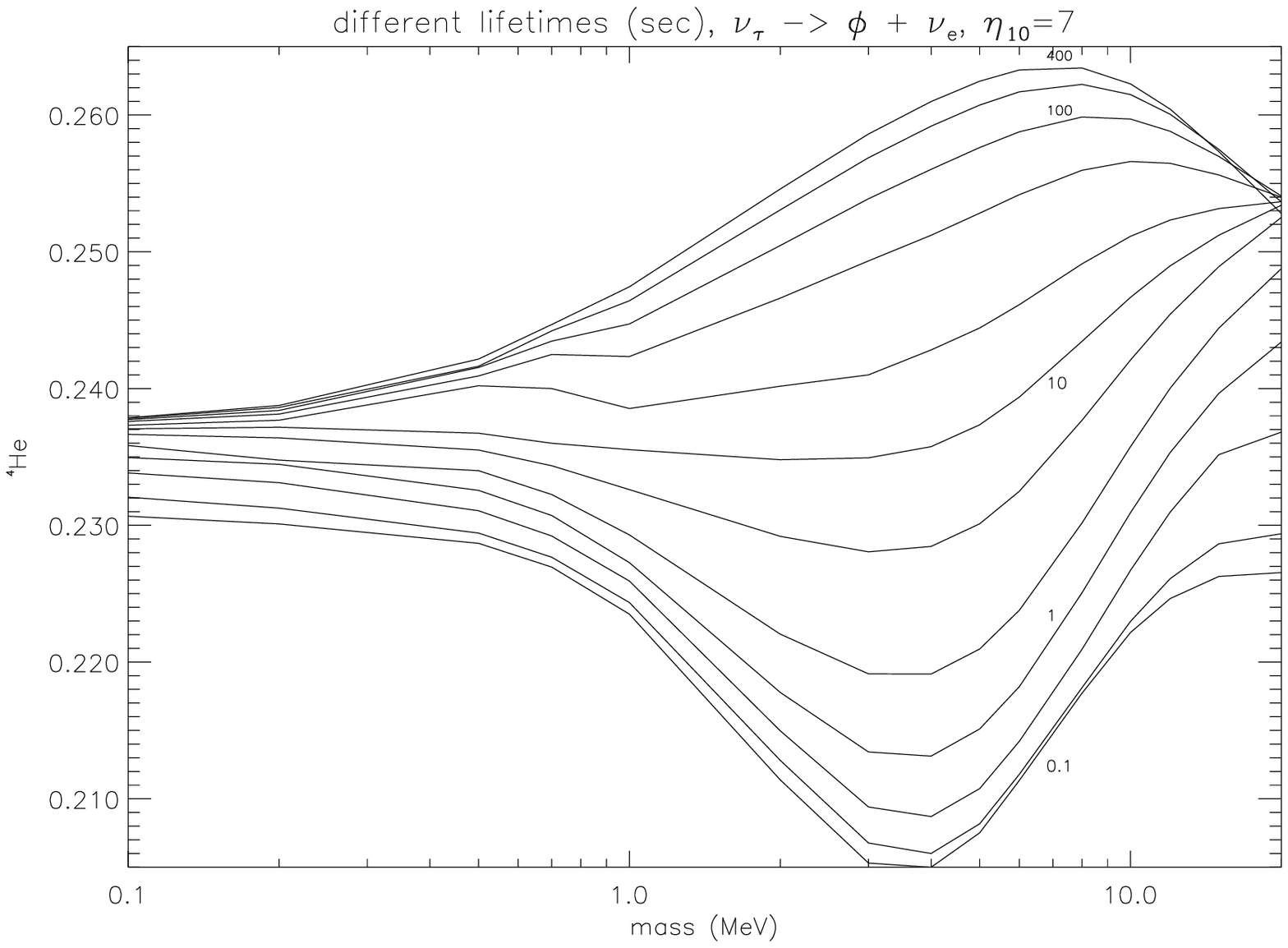,width=5in,height=3.5in}
\begin{center}
{\bf Figure 11a.}
\end{center}

\psfig{file=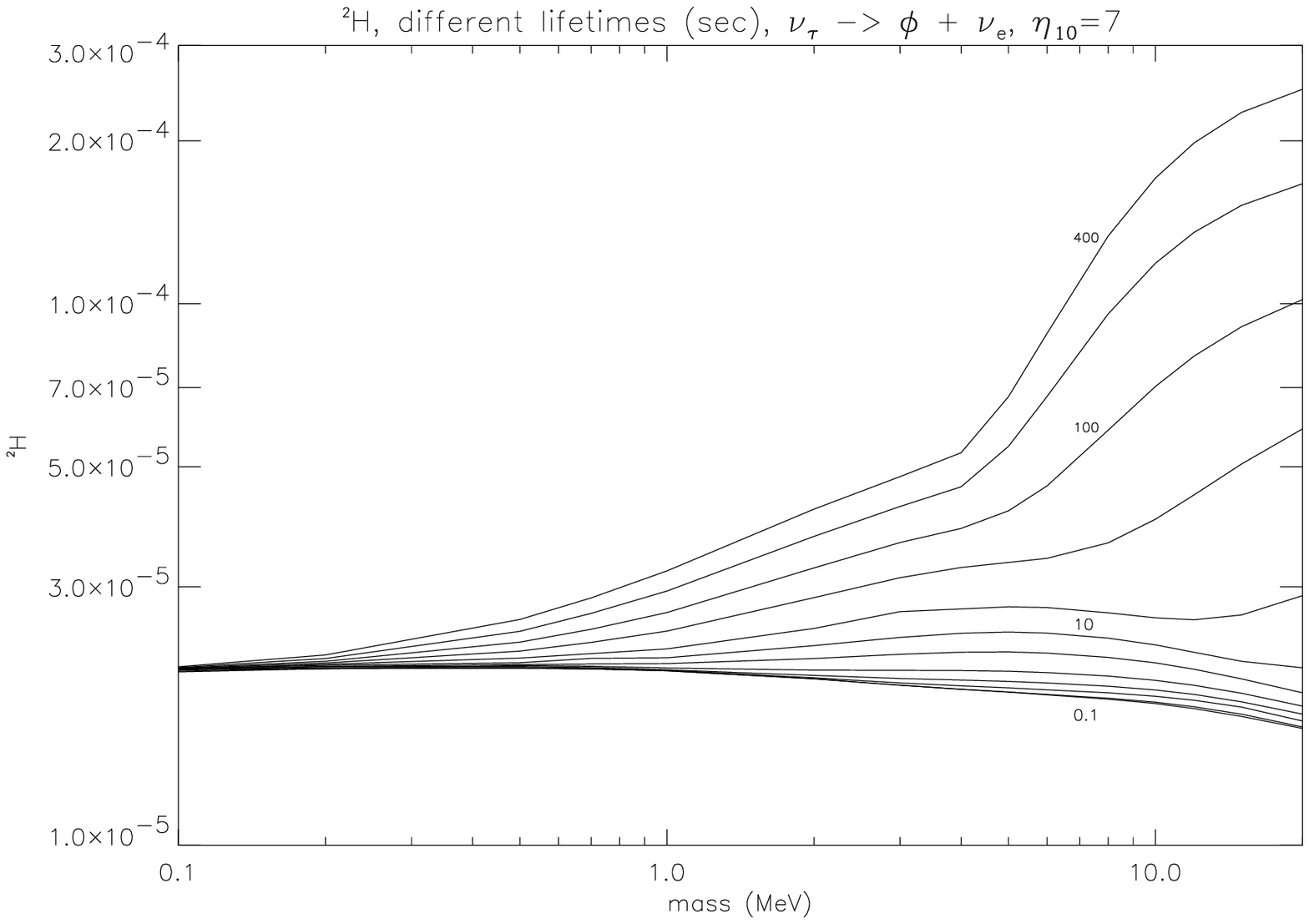,width=5in,height=3.5in}
\begin{center}
{\bf Figure 11b.}
\end{center}

\newpage

\psfig{file=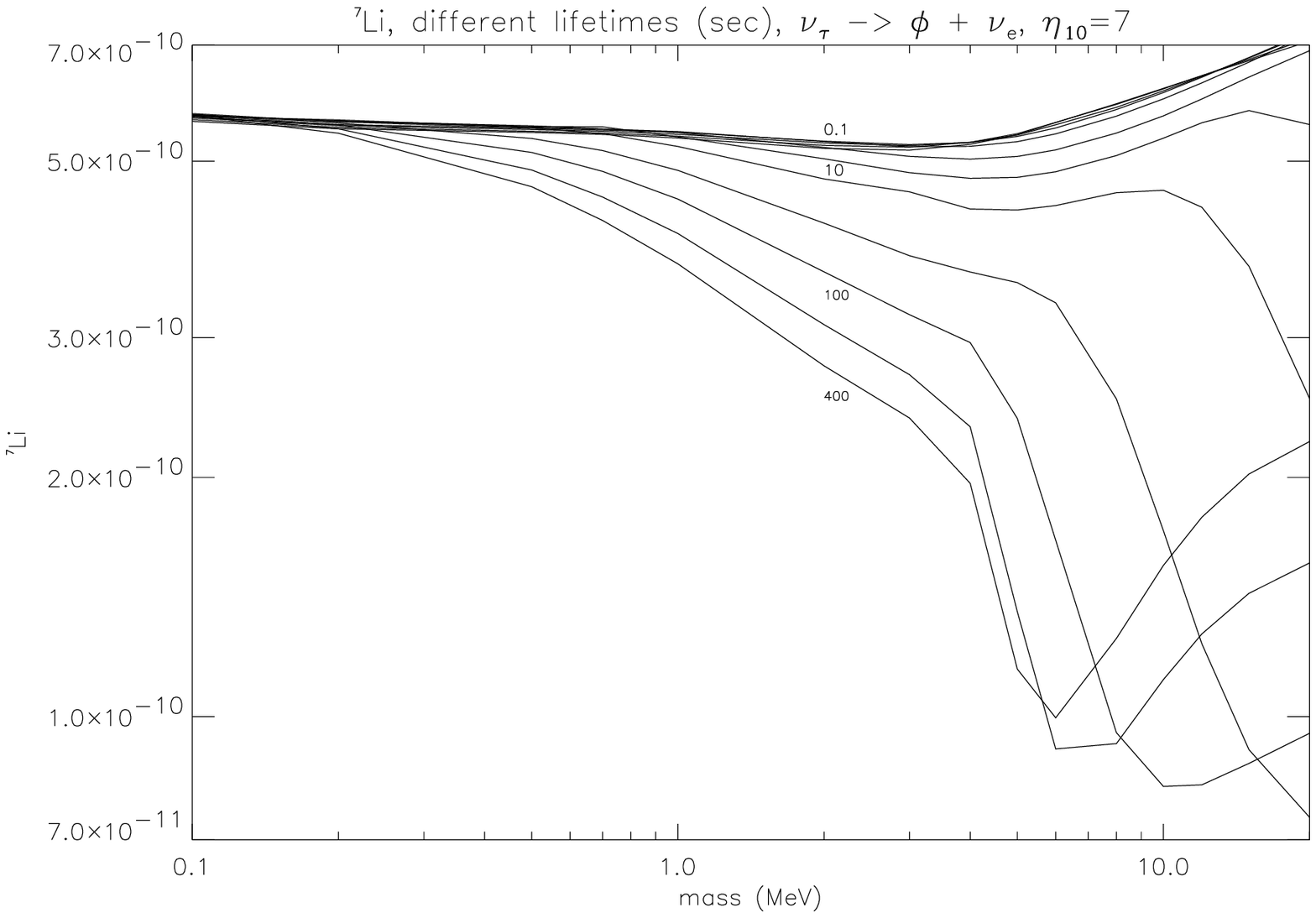,width=5in,height=3.5in}
\begin{center}
{\bf Figure 11c.}
\end{center}

\psfig{file=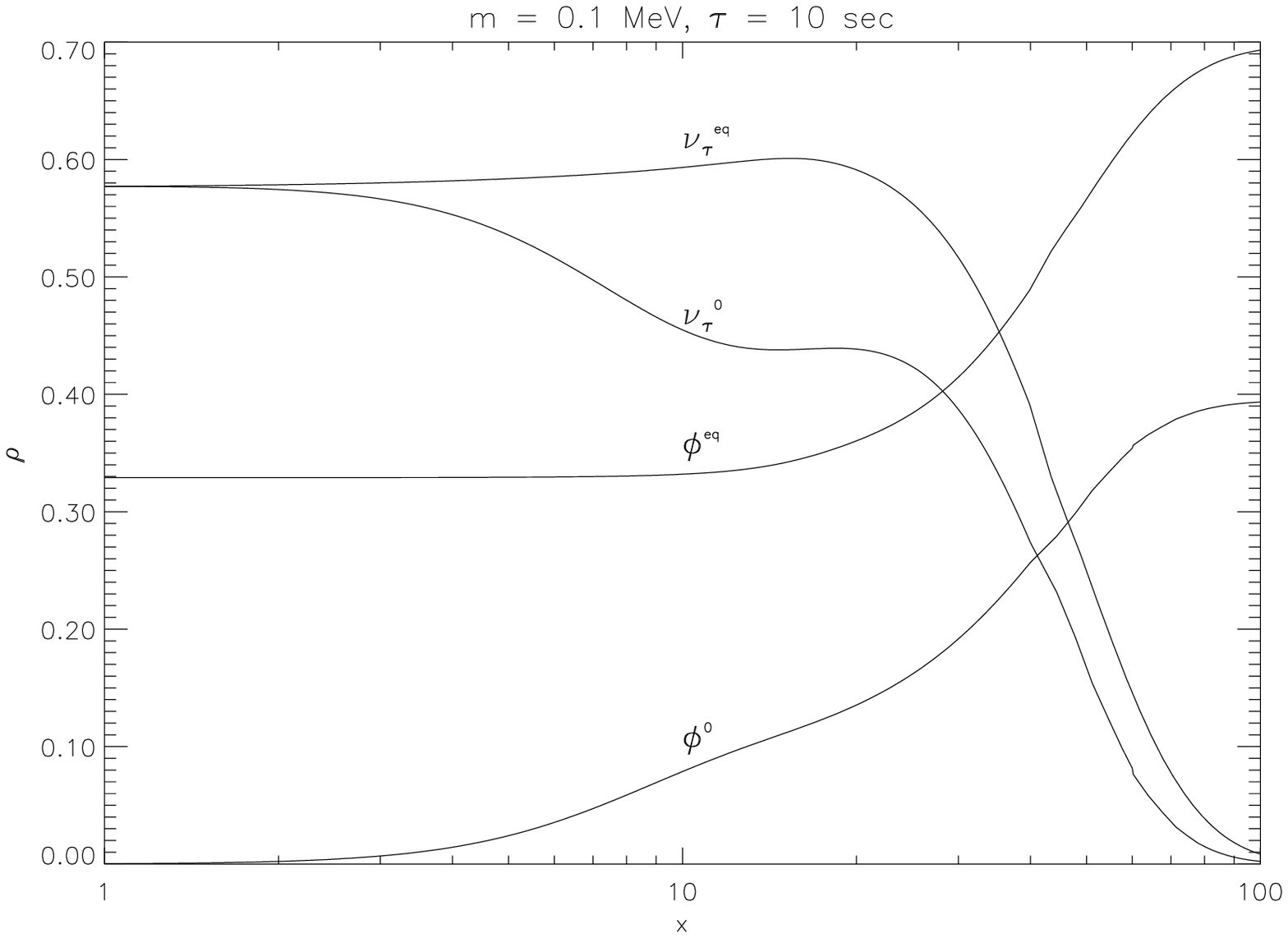,width=5in,height=3.5in}
\begin{center}
{\bf Figure 12.}
\end{center}

\end{document}